\documentclass{article}
\usepackage[utf8]{inputenc}
\usepackage{authblk}
\usepackage{graphicx}
\usepackage{caption}
\usepackage{subcaption}
\usepackage{cite}
\usepackage{multicol}
\usepackage{amsmath}
\usepackage{bm}
\usepackage{blindtext}
\usepackage[a4paper, total={8.0in, 10.5in}]{geometry}
\DeclareUnicodeCharacter{2212}{-}
\usepackage[T1, OT1]{fontenc}
\DeclareTextSymbolDefault{\dh}{T1}
\usepackage{hyperref}
\hypersetup{
       colorlinks = true,
       citecolor = blue,
       linkcolor = blue,
       urlcolor =  blue,
}

\title{\textbf{Josephson Transport across T-shaped and Series-Configured Double Quantum Dots System at Infinite-U Limit}}

\author{\bf Bhupendra Kumar \thanks{\bf bhupendra\_k@ph.iitr.ac.in}}
\author{Sachin Verma \thanks{sverma2@ph.iitr.ac.in}}
\author{Tanuj Chamoli \thanks{tchamoli@ph.iitr.ac.in}}

\author{Ajay \thanks{ajay@ph.iitr.ac.in}}
\affil{Department of Physics, Indian Institute of Technology Roorkee, 247667,
Uttarakhand, India}
\begin{document}

\maketitle

\begin{abstract}
The charge transport has been analyzed theoretically across a T-shaped and series-configured double quantum dots Josephson junction by implementing the Slave Boson mean field approximation at an infinite-U limit. It has been shown that Andreev Bound states (ABS) and Josephson current can be tuned by varying the interdot tunneling (t) and quantum dots energy level. For the T-shape configuration of the quantum dots, an extra path is available for the transport of electrons which causes the interference destruction between two paths. For decoupled quantum dots with $\epsilon_{d1}=\epsilon_{d2}=0$, the energy of ABS crosses at $\omega=0$ and Josephson current shows a discontinuity at $\phi=\pm \pi$. On the other hand, for coupled quantum dots the lower and upper ABS has a finite spacing, the Josephson current exhibits sinusoidal nature and its magnitude suppresses with increasing interdot tunneling strength. While in the series configuration, with increment in t, Josephson current increases and shows a discontinuity at $\phi=\pm \pi$, once the system gets resonant tunneling for $t=0.5\Gamma$ with $\epsilon_{d1}=\epsilon_{d2}=0.5\Gamma$. Further, we also analyze the nature of the energy of ABS and Josephson current with the quantum dots energy level in both configurations.

\end{abstract}

\begin{figure}[!ht]
  \begin{center}
    \includegraphics[width=0.5\textwidth]{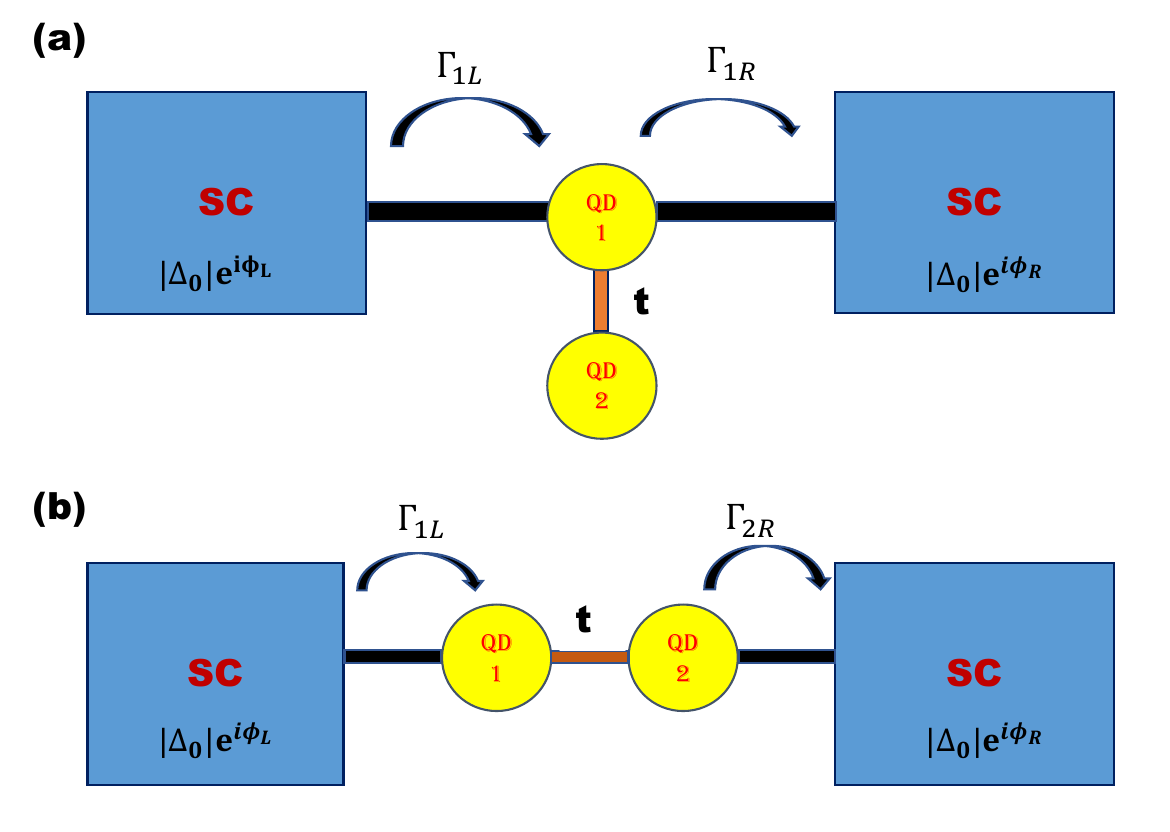}
  \end{center}
\caption{\small A schematic picture of a system with double quantum dots in (a) T-shape geometry and (b) series geometry connected to two superconducting leads with phase $\phi_{\alpha}$ where $\alpha \in L,R$.}
  \label{fig:Diagram_1}
\end{figure}

\begin{multicols}{2}
[
\begin{center}
\textbf{I. INTRODUCTION}
\end{center}
]
The Josephson effect is responsible for Cooper pair transport between two superconducting leads. It is the phase difference between two superconductors that lead to current flow; the sign of the current determines whether the state is called $0$-phase or $\pi$-phase \cite{Josephson1962,anderson1970josephson}. Josephson effect studies are interesting when an insulator is replaced by quantum dots in the junction, where Josephson current flows due to the formation of subgap states i.e. Andreev Bound states (ABS). The energy level of quantum dots can be tuned by the gate voltage \cite{kouwenhoven2001few,Kastner1993}. In a superconductor-quantum dot system, the Andreev reflection process occurs at the interfaces between the quantum dot and the superconducting leads. In this process, an electron from one of the leads strikes the dot, it can be retro-reflected as a hole, forming a Cooper pair in the superconducting lead. The existence of Andreev Bound states may significantly affect the system's transport characteristics. They have the potential to cause resonant tunneling as well as the creation of conductance peaks in the tunneling spectrum. Additionally, Andreev bound states can be utilized to design innovative quantum devices, such as qubits for quantum computing and superconducting nanoelectronics. A thorough overview of quantum dot-superconductor systems is provided in reference \cite{martin2011josephson,de2010hybrid}, which also includes studies of Andreev bound states and Josephson supercurrent in previous years \cite{PhysRevLett.99.126602,PhysRevLett.89.256801,sand2007kondo,r10,Cleuziou2006,PhysRevX.2.011009,r13,Verma2020,PhysRevLett.91.057005,r15,PhysRevLett.96.207003,doi:10.1021/nl071152w,PhysRevB.79.155441,Pillet2010,Hofstetter2009,PhysRevLett.104.026801,vecino2003josephson,choi2004kondo,lim2008andreev,zhu2001andreev,karrasch2008josephson}.
\\
It has been an active research topic in recent years to study quantum electronic transport across systems where double quantum dots are linked to superconducting leads, because of the tunable parameters such as energy levels of quantum dots, dot-lead coupling strength, and interdot tunneling. The interdot tunneling decides whether the dots are strongly or weakly coupled. These controlling tunable parameters of DQD make it helpful to analyze the transport phenomena like the Coulomb blockade, Kondo effect, quantum coherence, etc. A double quantum dot (DQD) system can be configured in three different ways: serially, T-shaped, or parallelly \cite{Cheng_2008, PhysRevB.75.045132,PhysRevB.107.115165}. There are number of theoretical studies addressing the Josephson current when coupled double quantum dots are connected to BCS superconducting lead at finite ondot Coulomb interactions (U) \cite{Chi2005,PhysRevB.66.085306,Droste2012, RAJPUT2014193,PhysRevLett.105.116803,r34}. In these theoretical studies, techniques such as non-equilibrium Green's function method \cite{haug2008quantum,keldysh1965diagram}, quantum Monte Carlo simulation \cite{PhysRevLett.108.227001}, perturbation theory \cite{Probst2016}, and renormalization group method have been used \cite{PhysRevB.92.014504,Wang2017,PhysRevB.62.13569,PhysRevB.89.235110,PhysRevB.95.045104,r51}. S-DQD-S systems have been formed through various experimental techniques \cite{Baba2017,PhysRevB.72.033414,RevModPhys.75.1,Nilsson2017,r39,PhysRevLett.128.046801}. The subgap gap states and Josephson current are also investigated at a very strong Coulomb correlation i.e. U is quite large in comparison to the other energy scales, $\Delta$, and interdot tunneling (t) and energy levels of quantum dots, in some research works \cite{r51,PhysRevB.67.041301,PhysRevLett.91.266802,Deacon2015}. Our assumption that ondot Coulomb repulsion is infinite is supported by this observation. Due to the infinite on-dot Coulomb interaction $U \to \infty$, both quantum dots, $QD_1$ and $QD_2$, are singly occupied. \\
In the present research work, we have applied Slave Boson mean-field approximation \cite{PhysRevB.75.045132,PhysRevB.29.3035,coleman1985large,PhysRevLett.58.266,PhysRevLett.57.1362,PhysRevB.59.1637,chamoli2022josephson,Chamoli2022} to study the Andreev Bound states and Josephson current through a T- shaped and series-configured double quantum dots Josephson junction. In T-shape configuration, $QD_1$ is directly coupled with superconducting leads, and $QD_2$ is only coupled with the $QD_1$ (figure \ref{fig:Diagram_1}a). In a series configuration, $QD_1$ is coupled with the left superconducting lead, and $QD_2$ is linked with the right superconducting lead (figure \ref{fig:Diagram_1}b). This infinite-U technique is sustainable for conditions where $T_K > \Delta$ where $T_K$ is the Kondo temperature and $\Delta$ is the Cooper pair binding energy \cite{PhysRevLett.110.076803,r62}.
\end{multicols}

\begin{multicols}{2}
[
\begin{center}
\textbf{II. THEORETICAL FORMULATION}
\end{center}
]
The double quantum dots system in T-shaped and in series configuration is modeled by generalized Anderson and BCS Hamiltonian in second quantization formalism as follows:
\\
\begin{center}
   \textbf{[A]: T-shaped configuration:}  
\end{center}

\begin{equation}
    \hat{H}=\hat{H}_{leads}+\hat{H}_{dots}+\hat{H}_{interdot-tunneling}+\hat{H}_{dot-lead}
\end{equation}
where 
\begin{equation*}
 \begin{aligned}
      \hat{H}_{leads} = \sum_{k\sigma,\alpha} \epsilon_{k\alpha}c^\dagger_{k\sigma,\alpha}c_{k\sigma,\alpha}-\left( \sum_{k\alpha}\Delta_{\alpha}c^\dagger_{k\uparrow,\alpha}c^\dagger_{-k\downarrow,\alpha}+h.c.)  \right)
 \end{aligned}
\end{equation*}
\begin{equation*}
\begin{aligned}
     \hat {H}_{dots}= & \sum_{i\sigma}\epsilon_{d_{i}}d^\dagger_{i\sigma}d_{i\sigma}
     +\sum_{i\sigma}U_{i}n_{i\sigma}n_{i\bar{\sigma}} +U_{12}n_{1}n_{2}
     \end{aligned}
\end{equation*}

\begin{equation*}
 \begin{aligned}
\hat{H}_{interdot-tunneling} = \sum_{i\sigma}t d^\dagger_{1\sigma}d_{2\sigma}+h.c
 \end{aligned}
\end{equation*}
\begin{equation*}
    \hat {H}_{dot-lead}=\sum_{k\sigma,\alpha}h_{1k,\alpha}c^\dagger_{k\sigma,\alpha}d_{1\sigma}+h.c
\end{equation*}
$\hat{H}_{leads}$ is the Hamiltonian of BCS superconducting leads. In the first term $\epsilon_{k\alpha}$ is the energy of superconducting leads ($\alpha \in L,R$) and ${c^\dagger}_{k\sigma,\alpha}(c_{k\sigma,\alpha})$ is the creation (annihilation) operator of electrons with spin $\sigma$ $(\uparrow,\downarrow)$ and wave vector $\vec{k}$. The second term denotes the interaction between Cooper pairs where $\Delta_{\alpha}$ is the superconducting order parameter and is given as $\Delta_{\alpha}=\lvert \Delta_0 \rvert e^{i\phi_{\alpha}}$.

$\hat {H}_{dots}$ denotes the Hamiltonian of double quantum dots. The energy of main quantum dot ($QD_1$) and side quantum dot ($QD_2$) is given by $\epsilon_{di}$ with fermionic operators ($d^\dagger_{i\sigma}$ and $d_{i\sigma}$).  $U_1$, $U_2$ are Coulomb interactions on $QD_1$ and $QD_2$ respectively. $U_{12}$ is the intra-dot Coulomb interaction.

$ \hat {H}_{interdot-tunneling}$ is the Hamiltonian for tunneling between both quantum dots and the amplitude of interdot tunneling is symbolized by t.

$ \hat {H}_{dot-lead}$ represents the hamiltonian for tunneling between $QD_{1}$ and superconducting leads where $h_{1k,\alpha}$ is the tunneling strength between leads and $QD_{1}$.

The Bogoliubov transformation is used to obtain the diagonalized Hamiltonian. 
\begin{equation}
    \begin{aligned}
c_{k\uparrow,\alpha}=u^*_{k\alpha}\beta_{k\uparrow,\alpha}+ v_{k\alpha}\beta^\dagger_{-k\downarrow,\alpha}
    \end{aligned}
\end{equation}
\begin{equation}
    \begin{aligned}
c_{k\downarrow,\alpha}=u^*_{k\alpha}\beta_{k\downarrow,\alpha}- v_{k\alpha}\beta^\dagger_{-k\uparrow,\alpha}
   \end{aligned}
\end{equation}
\begin{equation}
   \begin{aligned}
c^\dagger_{k\uparrow,\alpha}=u_{k\alpha}\beta^\dagger_{k\uparrow,\alpha}+ v^*_{k\alpha}\beta_{-k\downarrow,\alpha}
   \end{aligned}
\end{equation}
 \begin{equation}
c^\dagger_{k\downarrow,\alpha}=u_{k\alpha}\beta^\dagger_{k\downarrow,\alpha}- v_{k\alpha}^*\beta_{k\uparrow,\alpha}
 \end{equation}

After employing the above Bogoliubov transformation relations, the Hamiltonian can be expressed as:

\begin{equation}
    \begin{aligned}
        \hat{H}= & \sum_{k\alpha} E_{k\alpha}(\beta^\dagger_{k\uparrow,\alpha}\beta_{k\uparrow,\alpha}+\beta^\dagger_{-k\downarrow,\alpha}\beta_{-k\downarrow,\alpha})
        \\ &
        + \sum_{i\sigma}\epsilon_{di}d^\dagger_{i\sigma}d_{i\sigma}+\sum_{i\sigma}(td^\dagger_{1\sigma}d_{2\sigma}+t^*d^\dagger_{2\sigma}d_{1\sigma})
        \\ &
        + \sum_{i\sigma}U_{i}n_{i\sigma}n_{i\bar{\sigma}}+U_{12}n_1n_2
        \\ &
        + 
        \sum_{k\sigma}h_{1k,\alpha}[(u_{k\alpha}\beta^\dagger_{k\uparrow,\alpha}+v^*_{k\alpha}\beta_{-k\downarrow,\alpha})d_{1\uparrow}
        \\ &
        +
        (u_{k\alpha}\beta^\dagger_{k\uparrow,\alpha}-v^*_{k\alpha}\beta_{-k\uparrow,\alpha})d_{1\downarrow}]
        \\ &
        +
         \sum_{k\sigma}h^*_{1k,\alpha}[d^\dagger_{1\uparrow}(u_{k\alpha}^*\beta_{k\uparrow,\alpha}+v_{k\alpha}\beta^\dagger_{-k\downarrow,\alpha})
        \\ &
        +
        d^\downarrow_{1\downarrow}(u^*_{k\alpha}\beta_{k\downarrow,\alpha}-v_{k\alpha}\beta^\dagger_{-k\uparrow,\alpha})]      
    \end{aligned}
\end{equation}

where 
\begin{equation}
E_{k\alpha}=\sqrt{\epsilon^2_{k\alpha}+\lvert\Delta^2_{k\alpha}\rvert}
\end{equation}
\\
\begin{equation*}
    \lvert u^2_{k\alpha}\rvert=\frac{1}{2} \left(1+\frac{\epsilon_{k\alpha}}{\sqrt{\epsilon^2_{k\alpha}+\lvert\Delta^2_{k\alpha}}\rvert}\right)
\end{equation*}
\begin{equation}
 \lvert v^2_{k\alpha}\rvert=\frac{1}{2} \left(1-\frac{\epsilon_{k\alpha}}{\sqrt{\epsilon^2_{k\alpha}+\lvert\Delta^2_{k\alpha}}\rvert}\right)  
\end{equation}
\begin{equation*}
    u^*_{k\alpha}v_{k\alpha}=\frac{\Delta_{k\alpha}}{2E_{k\alpha}}
\end{equation*}
\begin{equation*}
    v^*_{k\alpha}u_{k\alpha}=\frac{\Delta^*_{k\alpha}}{2E_{k\alpha}}
\end{equation*}

To study the Andreev Bound states and supercurrent at infinite-U limit $(U_i \to \infty)$, we introduce the fermionic operators $f^\dagger_{i\sigma}$ and $f_{i\sigma}$in terms of bosonic operators $b^\dagger_{i}$ and $b_i$.
\begin{equation}
    d_{i\sigma}=b^\dagger_{i}f_{i\sigma}, 
    \\
    d^\dagger_{i\sigma}=f^\dagger_{i\sigma}b_i
\end{equation}

Here, $f_{i\sigma}$ denotes the singly occupied states and $b$ denotes the empty dot states. There are two quantum dots in our system therefore to apply the Infinite-U Slave Boson approximation approach, we have to introduce two Lagrange multipliers $\lambda_1$ and $\lambda_2$ to force the constraint in this fashion.
\begin{equation}
b^\dagger_{i}b_i+\sum_{\sigma}f^\dagger_{i\sigma}f_{i\sigma}=1
\end{equation}

For simplicity, we have neglected the intra-dot Coulomb repulsion $U_{12}$ because it is small as compared to other energy parameters. Finally, in the infinite-U limit, the effective Slave Boson mean field Hamiltonian with $\langle b^\dagger_{i}(t)\rangle=\langle b_{i}(t)\rangle=b_i$ can be expressed as:

\begin{equation}
        \begin{aligned}
        \hat{H}_{SBMF}= & \sum_{k\alpha} E_{k\alpha}(\beta^\dagger_{k\uparrow,\alpha}\beta_{k\uparrow,\alpha}+\beta^\dagger_{-k\downarrow,\alpha}\beta_{-k\downarrow,\alpha})
        \\ &
        + \sum_{i\sigma}\tilde{\epsilon}_{di}f^\dagger_{i\sigma}f_{i\sigma}+\sum_{i\sigma}(\tilde{t}f^\dagger_{1\sigma}f_{2\sigma}+\tilde{t}^*f^\dagger_{2\sigma}f_{1\sigma})
        \\ &
        + 
        \sum_{k\sigma}\tilde{h}_{1k,\alpha}[(u_{k\alpha}\beta^\dagger_{k\uparrow,\alpha}+v^*_{k\alpha}\beta_{-k\downarrow,\alpha})f_{1\uparrow}
        \\ &
        +
        (u_{k\alpha}\beta^\dagger_{k\uparrow,\alpha}-v^*_{k\alpha}\beta_{-k\uparrow,\alpha})f_{1\downarrow}]
        \\ &
        +
         \sum_{k\sigma}\tilde{h}^*_{1k,\alpha}[f^\dagger_{1\uparrow}(u_{k\alpha}^*\beta_{k\uparrow,\alpha}+v_{k\alpha}\beta^\dagger_{-k\downarrow,\alpha})
        \\ &
        +
        f^\downarrow_{1\downarrow}(u^*_{k\alpha}\beta_{k\downarrow,\alpha}-v_{k\alpha}\beta^\dagger_{-k\uparrow,\alpha})]
        \\ &
        +
        \lambda_1(b^2_{1}-1) + \lambda_2(b^2_{2}-1)
    \end{aligned}
\end{equation}
where $\tilde{\epsilon}_{di}$ are renormalized energy levels of quantum dots, $\Tilde{t}$ is the renormalized interdot tunneling strength, and $\tilde{h}_{1k,\alpha}$ is the renormalized dot lead tunneling strength, and can be written as:
\begin{equation*}
    \Tilde{h}_{1k,\alpha}=b_1h_{1k,\alpha}
\end{equation*}
\begin{equation}
    \Tilde{\epsilon}_{di}=\epsilon_{di}+\lambda_i
\end{equation}
\begin{equation*}
    \Tilde{t}=tb_{1}b_{2}
\end{equation*}

Based on Green's equation of motion technique, the following closed-coupled equations are obtained.
\begin{equation}
    \begin{aligned}
        (\omega-\tilde{\epsilon}_{d1})\langle\langle f_{1\uparrow}\vert f^\dagger_{1\uparrow}\rangle\rangle= & 1+ \sum_{k\alpha} \tilde{h}^*_{1k,\alpha}u^*_{k\alpha}\langle\langle \beta_{k\uparrow,\alpha}\vert f^\dagger_{1\uparrow}\rangle\rangle
    \\ & 
    + \sum_{k\alpha}\tilde{h}^*_{1k,\alpha}v_{k\alpha}\langle\langle \beta^\dagger_{-k\downarrow,\alpha}\vert f^\dagger_{1\uparrow}\rangle\rangle 
    \\ &
    + \tilde{t} \langle\langle f_{2\uparrow}\vert f^\dagger_{1\uparrow}\rangle\rangle 
    \end{aligned}
\end{equation}
\begin{equation}
\begin{aligned}
(\omega-E_{k\alpha})\langle\langle \beta_{k\uparrow,\alpha}\vert f^\dagger_{1\uparrow}\rangle\rangle = & \sum_{k\alpha} \tilde{h}_{1k,\alpha}u_{k}\langle\langle f_{1\uparrow}\vert f^\dagger_{1\uparrow}\rangle\rangle
\\ &
    + \sum_{k\alpha}\tilde{h}^*_{1k,\alpha}v_{k\alpha}\langle\langle f^\dagger_{1\downarrow}\vert f^\dagger_{1\uparrow}\rangle\rangle
\end{aligned}
\end{equation}
\begin{equation}
\begin{aligned}
    (\omega + E_{k\alpha})\langle\langle \beta^\dagger_{-k\downarrow,\alpha}\vert f^\dagger_{1\uparrow}\rangle\rangle = & \sum_{k\alpha} \tilde{h}_{1k,\alpha} v^*_{k\alpha} \langle\langle f_{1\uparrow}\vert f^\dagger_{1\uparrow}\rangle\rangle
  \\ &
    -\sum_{k\alpha} \tilde{h}^*_{1k,\alpha} u^*_{k\alpha} \langle\langle f^\dagger_{1\downarrow}\vert f^\dagger_{1\uparrow}\rangle\rangle
\end{aligned} 
\end{equation}
\begin{equation}
\begin{aligned}
(\omega + \tilde{\epsilon}_{d1})\langle\langle f^\dagger_{1\downarrow}\vert f^\dagger_{1\uparrow}\rangle\rangle  = & \sum_{k\alpha} \tilde{h}_{1k,\alpha} v^*_{\alpha}  \langle\langle \beta_{-k\uparrow,\alpha}\vert f^\dagger_{1\uparrow}  \rangle\rangle
\\ &
   - \sum_{k\alpha} \tilde{h}_{1k,\alpha}u_{k\alpha} \langle\langle \beta^\dagger_{k\downarrow,\alpha}\vert f^\dagger_{1\uparrow}\rangle\rangle 
\\ &
   - \tilde{t}^* \langle\langle f^\dagger_{2\downarrow}\vert f^\dagger_{1\uparrow}\rangle\rangle
\end{aligned}
\end{equation}
\begin{equation}
\begin{aligned}
   (\omega-E_{k\alpha})\langle\langle \beta_{-k\uparrow,\alpha}\vert f^\dagger_{1\uparrow}\rangle\rangle = & \sum_{k\alpha} \tilde{h}_{1k,\alpha}u_{k\alpha}\langle\langle f_{1\uparrow}\vert f^\dagger_{1\uparrow}\rangle\rangle
\\ &
    + \sum_{k\alpha}\tilde{h}^*_{1k,\alpha}v_{k\alpha}\langle\langle f^\dagger_{1\downarrow}\vert f^\dagger_{1\uparrow}\rangle\rangle
\end{aligned} 
\end{equation}
\begin{equation}
\begin{aligned}
    (\omega + E_{k\alpha})\langle\langle \beta^\dagger_{k\downarrow,\alpha}\vert f^\dagger_{1\uparrow}\rangle\rangle = & \sum_{k\alpha} \tilde{h}_{1k,\alpha} v^*_{k\alpha} \langle\langle f_{1\uparrow}\vert f^\dagger_{1\uparrow}\rangle\rangle
  \\ &
    -\sum_{k\alpha} \tilde{h}^*_{1k,\alpha} u^*_{k\alpha} \langle\langle f^\dagger_{1\downarrow}\vert f^\dagger_{1\uparrow}\rangle\rangle
\end{aligned} 
\end{equation}
\begin{equation}
    \begin{aligned}
      (\omega-\tilde{\epsilon}_{d2})\langle\langle f_{1\uparrow}\vert f^\dagger_{1\uparrow}\rangle\rangle= & \tilde{t}^*\langle\langle f_{1\uparrow}\vert f^\dagger_{1\uparrow}\rangle\rangle
    \end{aligned}
\end{equation}
\begin{equation}
    \begin{aligned}
      (\omega+\tilde{\epsilon}_{d2})\langle\langle f_{1\uparrow}\vert f^\dagger_{1\uparrow}\rangle\rangle= & -\tilde{t}\langle\langle f^\dagger_{1\downarrow}\vert f^\dagger_{1\uparrow}\rangle\rangle
    \end{aligned}
\end{equation}
After solving the set of closed-coupled equations (Eq.13-20), the single particle retarded Green's function for $QD$ can be written as:
\begin{equation}
    \begin{aligned}
        G_{11}=\langle\langle f_{1\uparrow}\vert f^\dagger_{1\uparrow} \rangle\rangle = & \frac{1}{\omega-\tilde{\epsilon}_{d1}-\tilde{\zeta}_{2\alpha}-\frac{\tilde{t}^2}{\omega-\tilde{\epsilon}_{d2}}-\frac{\tilde{\zeta}_{3\alpha}\tilde{\zeta}_{4\alpha}}{\omega+\Tilde{\epsilon}_{d1}-\frac{\tilde{t}^2}{\omega+\tilde{\epsilon}_{d2}}-\tilde{\zeta}_{1\alpha}}}
        \\ &
        = 
        \frac{1}{\omega-\Tilde{\epsilon}_{d1}-\Tilde{\Lambda}}
    \end{aligned}
\end{equation}
where
\begin{equation}
    \begin{aligned}
      \Tilde{\Lambda}=\tilde{\zeta}_{2\alpha}+\frac {\tilde{t}^2}{\omega-\tilde{\epsilon}_{d2}}+\frac{\tilde{\zeta}_{3\alpha}\tilde{\zeta}_{4\alpha}}{\omega+\Tilde{\epsilon}_{d1}-\frac{\tilde{t}^2}{\omega+\tilde{\epsilon}_{d2}}-\tilde{\zeta}_{1\alpha}}
    \end{aligned}
\end{equation}
and also
\begin{equation}
    \begin{aligned}
      \tilde{\zeta}_{1\alpha}= & \sum_{k\alpha} \lvert \Tilde{h}_{1k,\alpha} \rvert^2 \left(\frac{\lvert u_{k\alpha}\rvert^2}{\omega+E_{k\alpha}}+\frac{\lvert v_{k\alpha}\rvert^2}{\omega-E_{k\alpha}} \right)
      \\ &
      =
      -i\frac{\tilde{\Gamma}_{1\alpha}}{2}\gamma_{\alpha}(\omega)
    \end{aligned}
\end{equation}
\begin{equation}
    \begin{aligned}
      \tilde{\zeta}_{2\alpha}= & \sum_{k\alpha} \lvert \Tilde{h}_{1k,\alpha} \rvert^2 \left(\frac{\lvert u_{k\alpha}\rvert^2}{\omega-E_{k\alpha}}+\frac{\lvert v_{k\alpha}\rvert^2}{\omega+E_{k\alpha}} \right)
      \\ &
      =
      -i\frac{\tilde{\Gamma}_{1\alpha}}{2}\gamma_{\alpha}(\omega)
    \end{aligned}
\end{equation}
\begin{equation}
    \begin{aligned}
      \tilde{\zeta}_{3\alpha}= & \sum_{k\alpha} \lvert \Tilde{h}_{1k,\alpha} \rvert^2 v^*_{k\alpha}u_{k\alpha} \left(\frac{1}{\omega-E_{k\alpha}}-\frac{1}{\omega+E_{k\alpha}} \right)
      \\ &
      =
      -i\frac{\tilde{\Gamma}_{1\alpha}}{2}\gamma_{\alpha}(\omega)\frac{\Delta^*}{\omega}
    \end{aligned}
\end{equation}
\begin{equation}
    \begin{aligned}
      \tilde{\zeta}_{4\alpha}= & \sum_{k\alpha} \lvert \Tilde{h}_{1k,\alpha} \rvert^2 u^*_{k\alpha}v_{k\alpha} \left(\frac{1}{\omega-E_{k\alpha}}-\frac{1}{\omega+E_{k\alpha}} \right)
      \\ &
      =
      -i\frac{\tilde{\Gamma}_{1\alpha}}{2}\gamma_{\alpha}(\omega)\frac{\Delta}{\omega}
    \end{aligned}
\end{equation}
where $\tilde{\Gamma}_{1\alpha}$ represents the renormalized coupling strength between $QD_1$ and superconducting leads and defines as $\tilde{\Gamma}_{1\alpha}=\lvert b_{i}\rvert^2 \Gamma_{1\alpha}$. $\Gamma_{1\alpha}=2\pi \gamma_{\alpha}\lvert h_{1\alpha}\rvert^2$ stands for coupling strength. The BCS superconducting density of states $(\gamma_{\alpha})$ can be written as follows.
\begin{equation}
    \begin{aligned}
        \gamma_{\alpha}(\omega)=\frac{\omega \theta(\Delta_{\alpha}-\lvert \omega \rvert)}{\sqrt{\Delta^2_{\alpha}-\omega^2}}+i\frac{\lvert \omega \rvert \theta(\lvert \omega \rvert-\Delta_{\alpha})}{\sqrt{\omega^2-\Delta_{\alpha}^2}}
    \end{aligned}
\end{equation}

\begin{center}
 \textbf{[B]:  Series configuration: }  
\end{center}

\begin{equation}  \hat{H}=\hat{H}_{leads}+\hat{H}_{dots}+\hat{H}_{interdot-tunneling}+\hat{H}_{dot-lead}
\end{equation}
where 
\begin{equation*}
 \begin{aligned}
      \hat{H}_{leads} = & \sum_{k\sigma} \epsilon_{kL}c^\dagger_{k\sigma,L}c_{k\sigma,L} + \sum_{k\sigma} \epsilon_{kR}c^\dagger_{k\sigma,R}c_{k\sigma,R}
      \\ &
      -\left( \sum_{k\alpha}\Delta_{\alpha}c^\dagger_{k\uparrow,\alpha}c^\dagger_{-k\downarrow,\alpha}+h.c.  \right)    
 \end{aligned}
\end{equation*}
\begin{equation*}
\begin{aligned}
     \hat {H}_{dots}= & \sum_{i\sigma}\epsilon_{d_{i}}d^\dagger_{i\sigma}d_{i\sigma}
+\sum_{i\sigma}U_{i}n_{i\sigma}n_{i\bar{\sigma}} +U_{12}n_{1}n_{2}
     \end{aligned}
\end{equation*}

\begin{equation*}
\begin{aligned}
     \hat{H}_{interdot-tunneling}= \sum_{i\sigma}t d^\dagger_{1\sigma}d_{2\sigma}+h.c
     \end{aligned}
\end{equation*}
\begin{equation*}
    \hat {H}_{dot-lead}=\sum_{k\sigma} \left (h_{1k,L}c^\dagger_{k\sigma,\alpha}d_{1\sigma} +h_{2k,R}c^\dagger_{k\sigma,R}d_{2\sigma} \right )+h.c
\end{equation*}

Again, by employing the Bogoliubov transformation and introducing the fermionic operator, the effective Slave Boson mean field Hamiltonian can be expressed as:
\begin{equation}
        \begin{aligned}
        \hat{H}_{SBMF}= & \sum_{k} E_{kL}(\beta^\dagger_{k\uparrow,L}\beta_{k\uparrow,L}+\beta^\dagger_{-k\downarrow,L}\beta_{-k\downarrow,L})
        \\ &
        +
        \sum_{k} E_{kR}(\beta^\dagger_{k\uparrow,R}\beta_{k\uparrow,R}+\beta^\dagger_{-k\downarrow,R}\beta_{-k\downarrow,R})
        \\ &
        + \sum_{\sigma}\tilde{\epsilon}_{d1}f^\dagger_{1\sigma}f_{1\sigma} +\sum_{\sigma}\tilde{\epsilon}_{d2}f^\dagger_{2\sigma}f_{2\sigma}
        \\ &
        +(\tilde{t}f^\dagger_{1\sigma}f_{2\sigma}+\tilde{t}^*f^\dagger_{2\sigma}f_{1\sigma})
        \\ &
        + 
        \sum_{k\sigma}\tilde{h}_{1k,L}[(u_{kL}\beta^\dagger_{k\uparrow,L}+v^*_{kL}\beta_{-k\downarrow,L})f_{1\uparrow}
        \\ &
        +
        (u_{kL}\beta^\dagger_{k\uparrow,L}-v^*_{kL}\beta_{-k\uparrow,L})f_{1\downarrow}]
        \\ &
        +
         \sum_{k\sigma}\tilde{h}^*_{1k,L}[f^\dagger_{1\uparrow}(u_{kL}^*\beta_{k\uparrow,L}+v_{kL}\beta^\dagger_{-k\downarrow,L})
        \\ &
        +
        f^\dagger_{1\downarrow}(u^*_{kL}\beta_{k\downarrow,L}-v_{kL}\beta^\dagger_{-k\uparrow,L})]
        \\ &
        +
         \sum_{k\sigma}\tilde{h}_{2k,R}[(u_{kR}\beta^\dagger_{k\uparrow,R}+v^*_{kR}\beta_{-k\downarrow,R})f_{1\uparrow}
        \\ &
        +
        (u_{kR}\beta^\dagger_{k\uparrow,R}-v^*_{kR}\beta_{-k\uparrow,R})f_{1\downarrow}]
        \\ &
        +
         \sum_{k\sigma}\tilde{h}^*_{2k,R}[f^\dagger_{1\uparrow}(u_{kR}^*\beta_{k\uparrow,R}+v_{kR}\beta^\dagger_{-k\downarrow,R})
        \\ &
        +
        f^\dagger_{1\downarrow}(u^*_{kR}\beta_{k\downarrow,R}-v_{kR}\beta^\dagger_{-k\uparrow,R})]
        \\ &
        +
        \lambda_1(b^2_{1}-1) + \lambda_2(b^2_{2}-1)
    \end{aligned}
\end{equation}

Based on Green's equation of motion technique, the following closed-coupled equations are obtained for series configured double quantum dots coupled to superconducting leads.
\begin{equation}
    \begin{aligned}
        (\omega-\tilde{\epsilon}_{d1})\langle\langle f_{1\uparrow}\vert f^\dagger_{1\uparrow}\rangle\rangle= & 1+ \sum_{k} \tilde{h}^*_{1k,L}u^*_{kL}\langle\langle \beta_{k\uparrow,L}\vert f^\dagger_{1\uparrow}\rangle\rangle
    \\ & 
    + \sum_{k}\tilde{h}^*_{1k,L}v_{kL}\langle\langle \beta^\dagger_{-k\downarrow,L}\vert f^\dagger_{1\uparrow}\rangle\rangle 
    \\ &
    + \tilde{t} \langle\langle f_{2\uparrow}\vert f^\dagger_{1\uparrow}\rangle\rangle 
    \end{aligned}
\end{equation}
\begin{equation}
\begin{aligned}
(\omega-E_{kL})\langle\langle \beta_{k\uparrow,L}\vert f^\dagger_{1\uparrow}\rangle\rangle = & \sum_{k} \tilde{h}_{1k,L}u_{k}\langle\langle f_{1\uparrow}\vert f^\dagger_{1\uparrow}\rangle\rangle
\\ &
    + \sum_{k}\tilde{h}^*_{1k,L}v_{kL}\langle\langle f^\dagger_{1\downarrow}\vert f^\dagger_{1\uparrow}\rangle\rangle
\end{aligned}
\end{equation}
\begin{equation}
\begin{aligned}
    (\omega + E_{kL})\langle\langle \beta^\dagger_{-k\downarrow,L}\vert f^\dagger_{1\uparrow}\rangle\rangle = & \sum_{k} \tilde{h}_{1k,L} v^*_{kL} \langle\langle f_{1\uparrow}\vert f^\dagger_{1\uparrow}\rangle\rangle
  \\ &
    -\sum_{k} \tilde{h}^*_{1k,L} u^*_{kL} \langle\langle f^\dagger_{1\downarrow}\vert f^\dagger_{1\uparrow}\rangle\rangle
\end{aligned} 
\end{equation}
\begin{equation}
\begin{aligned}
(\omega + \tilde{\epsilon}_{d1})\langle\langle f^\dagger_{1\downarrow}\vert f^\dagger_{1\uparrow}\rangle\rangle  = & \sum_{k} \tilde{h}_{1k,L} v^*_{kL}  \langle\langle \beta_{-k\uparrow,L}\vert f^\dagger_{1\uparrow}  \rangle\rangle
\\ &
   - \sum_{k} \tilde{h}_{1k,L}u_{kL} \langle\langle \beta^\dagger_{k\downarrow,L}\vert f^\dagger_{1\uparrow}\rangle\rangle 
\\ &
   - \tilde{t}^* \langle\langle f^\dagger_{2\downarrow}\vert f^\dagger_{1\uparrow}\rangle\rangle
\end{aligned}
\end{equation}
\begin{equation}
\begin{aligned}
   (\omega-E_{kR})\langle\langle \beta_{k\uparrow,R}\vert f^\dagger_{1\uparrow}\rangle\rangle = & \sum_{k} \tilde{h}_{2k,R}u_{k\alpha}\langle\langle f_{1\uparrow}\vert f^\dagger_{1\uparrow}\rangle\rangle
\\ &
    + \sum_{k\alpha}\tilde{h}^*_{2k,R}v_{kR}\langle\langle f^\dagger_{1\downarrow}\vert f^\dagger_{1\uparrow}\rangle\rangle
\end{aligned} 
\end{equation}
\begin{equation}
\begin{aligned}
    (\omega + E_{kR})\langle\langle \beta^\dagger_{-k\downarrow,R}\vert f^\dagger_{1\uparrow}\rangle\rangle = & \sum_{k} \tilde{h}_{2k,R} v^*_{kR} \langle\langle f_{1\uparrow}\vert f^\dagger_{1\uparrow}\rangle\rangle
  \\ &
    -\sum_{k} \tilde{h}^*_{2k,R} u^*_{kR} \langle\langle f^\dagger_{1\downarrow}\vert f^\dagger_{1\uparrow}\rangle\rangle
\end{aligned} 
\end{equation}
\begin{equation}
    \begin{aligned}
      (\omega-\tilde{\epsilon}_{d2})\langle\langle f_{2\uparrow}\vert f^\dagger_{1\uparrow}\rangle\rangle= & t\langle\langle f_{1\uparrow}\vert f^\dagger_{1\uparrow}\rangle\rangle
      \\ & 
      +\tilde{h}_{2k,R}u^*_{kR}\langle\langle \beta_{k\uparrow,R}f^\dagger_{1\uparrow}
      \\ &
      +\tilde{h}^*_{2k,R}v_{kR}\langle\langle
      \beta^\dagger_{-k\downarrow,R}f^\dagger_{1\uparrow}
    \end{aligned}
\end{equation}
\begin{equation}
    \begin{aligned}
      (\omega+\tilde{\epsilon}_{d2})\langle\langle f_{1\uparrow}\vert f^\dagger_{1\uparrow}\rangle\rangle=  & -t\langle\langle f^\dagger_{1\downarrow}\vert f^\dagger_{1\uparrow}\rangle\rangle
      \\ & 
      +\tilde{h}_{2k,R}v^*_{kR}\langle\langle \beta_{-k\uparrow,R}f^\dagger_{1\uparrow}
      \\ &
      -\tilde{h}_{2k,R}u_{kR}\langle\langle
      \beta^\dagger_{k\downarrow,R}f^\dagger_{1\uparrow}
    \end{aligned}
\end{equation}
After solving the set of closed-coupled equations (Eq.30-Eq.37), the single particle retarded Green's function for $QD$ can be written as:

\begin{equation}
\begin{aligned}
    G_{11}=\frac{1}{\omega-\tilde{\epsilon}_{d1}-\Tilde{\Lambda}}
\end{aligned}
\end{equation}

\begin{equation*}
\begin{aligned}
\Tilde{\Lambda}= A-\frac{B\cdot C}{D}
\end{aligned}
\end{equation*}
where 
\begin{equation*}
\begin{aligned}
A=\tilde{\zeta}_{1L}(2)+\frac{\tilde{t}^2}{\omega-\tilde{\epsilon}_{d2}-\tilde{\zeta}_{2R}(2)-\frac{\tilde{\zeta}_{3R}\tilde{\zeta}_{4R}}{\omega+\tilde{\epsilon}_{d2}-\tilde{\zeta}_{2R}(1)}}
\end{aligned}
\end{equation*}
\begin{equation*}
\begin{aligned}
B=\tilde{\zeta}_{4L}-\frac{\tilde{t}^2\cdot \tilde{\zeta}_{4R}}{(\omega+\tilde{\epsilon}_{d2}-\tilde{\zeta}_{2R}(1))(\omega-\tilde{\epsilon}_{d2}-\tilde{\zeta}_{2R}(2)-\frac{\tilde{\zeta}_{3R}\tilde{\zeta}_{4R}}{\omega+\tilde{\epsilon}_{d2}-\tilde{\zeta}_{2R}(1)})}
\end{aligned}
\end{equation*}

\begin{equation*}
\begin{aligned}
C=\tilde{\zeta}_{3L}-\frac{\tilde{t}^2\cdot \tilde{\zeta}_{3R}}{(\omega-\tilde{\epsilon}_{d2}-\tilde{\zeta}_{2R}(2))(\omega+\tilde{\epsilon}_{d2}-\tilde{\zeta}_{2R}(1)-\frac{\tilde{\zeta}_{3R}\tilde{\zeta}_{4R}}{\omega-\tilde{\epsilon}_{d2}-\tilde{\zeta}_{2R}(2)})}
\end{aligned}
\end{equation*}

\begin{equation*}
\begin{aligned}
D= \omega+\tilde{\epsilon}_{d1}-\tilde{\zeta}_{1L}(1)-\frac{\tilde{t}^2}{\omega+\tilde{\epsilon}_{d2}-\tilde{\zeta}_{2R}(1)-\frac{\zeta_{3R}\tilde{\zeta}_{4R}}{\omega-\tilde{\epsilon}_{d2}-\tilde{\zeta}_{2R}(2)}}
\end{aligned}
\end{equation*}
Also, where
\begin{equation}
    \begin{aligned}
      \tilde{\zeta}_{1L}(1)= & \sum_{k} \lvert \Tilde{h}_{1k,L} \rvert^2 \left(\frac{\lvert u_{kL}\rvert^2}{\omega+E_{kL}}+\frac{\lvert v_{kL}\rvert^2}{\omega-E_{kL}} \right)
      \\ &
      =
      -i\frac{\tilde{\Gamma}_{1L}}{2}\gamma_{L}(\omega)
    \end{aligned}
\end{equation}

\begin{equation}
    \begin{aligned}
      \tilde{\zeta}_{1L}(2)= & \sum_{k} \lvert \Tilde{h}_{1k,L} \rvert^2 \left(\frac{\lvert u_{kL}\rvert^2}{\omega-E_{kL}}+\frac{\lvert v_{kL}\rvert^2}{\omega+E_{kL}} \right)
      \\ &
      =
      -i\frac{\tilde{\Gamma}_{1L}}{2}\gamma_{L}(\omega)
    \end{aligned}
\end{equation}
\begin{equation}
    \begin{aligned}
      \tilde{\zeta}_{2R}(1)= & \sum_{k} \lvert \Tilde{h}_{2k,R} \rvert^2 \left(\frac{\lvert u_{kR}\rvert^2}{\omega+E_{kR}}+\frac{\lvert v_{kR}\rvert^2}{\omega-E_{kR}} \right)
      \\ &
      =
      -i\frac{\tilde{\Gamma}_{2R}}{2}\gamma_{R}(\omega)
    \end{aligned}
\end{equation}

\begin{equation}
    \begin{aligned}
      \tilde{\zeta}_{2R}(2)= & \sum_{k} \lvert \Tilde{h}_{2k,R} \rvert^2 \left(\frac{\lvert u_{kR}\rvert^2}{\omega-E_{kR}}+\frac{\lvert v_{kR}\rvert^2}{\omega+E_{kR}} \right)
      \\ &
      =
      -i\frac{\tilde{\Gamma}_{2R}}{2}\gamma_{R}(\omega)
    \end{aligned}
\end{equation}

\begin{equation}
    \begin{aligned}
      \tilde{\zeta}_{3L}= & \sum_{k} \lvert \Tilde{h}_{1k,L} \rvert^2 v^*_{kL}u_{kL} \left(\frac{1}{\omega-E_{kL}}-\frac{1}{\omega+E_{kL}} \right)
      \\ &
      =
      -i\frac{\tilde{\Gamma}_{1L}}{2}\gamma_{L}(\omega)\frac{\Delta^*}{\omega}
    \end{aligned}
\end{equation}

\begin{equation}
    \begin{aligned}
      \tilde{\zeta}_{3R}= & \sum_{k} \lvert \Tilde{h}_{2k,R} \rvert^2 v^*_{kR}u_{kR} \left(\frac{1}{\omega-E_{kR}}-\frac{1}{\omega+E_{kR}} \right)
      \\ &
      =
      -i\frac{\tilde{\Gamma}_{2R}}{2}\gamma_{L}(\omega)\frac{\Delta^*}{\omega}
    \end{aligned}
\end{equation}

\begin{equation}
    \begin{aligned}
      \tilde{\zeta}_{4L}= & \sum_{k} \lvert \Tilde{h}_{1k,L} \rvert^2 v_{kL}u^*_{kL} \left(\frac{1}{\omega-E_{kL}}-\frac{1}{\omega+E_{kL}} \right)
      \\ &
      =
      -i\frac{\tilde{\Gamma}_{1L}}{2}\gamma_{L}(\omega)\frac{\Delta}{\omega}
    \end{aligned}
\end{equation}

\begin{equation}
    \begin{aligned}
      \tilde{\zeta}_{4R}= & \sum_{k} \lvert \Tilde{h}_{2k,R} \rvert^2 v_{kR}u^*_{kR} \left(\frac{1}{\omega-E_{kR}}-\frac{1}{\omega+E_{kR}} \right)
      \\ &
      =
      -i\frac{\tilde{\Gamma}_{2R}}{2}\gamma_{R}(\omega)\frac{\Delta}{\omega}
    \end{aligned}
\end{equation}
where $\tilde{\Gamma}_{1L}$ and $\tilde{\Gamma}_{2R}$ represents the renormalized coupling strength between $QD_1$ with left superconducting lead and $QD_2$ with right superconducting lead respectively.

The mean free energy for T-shaped and series-configured double quantum dot system can be calculated by using the following formula \cite{coleman1985large}: 
\begin{equation}
    \begin{aligned}
        F_{MF}= & -\frac{N}{\pi}\int f(\omega) \tan^{-1}\left( \frac{\tilde{\Lambda}}{\tilde{\epsilon}_{d1}-\omega}\right)
        \\ & 
        + \lambda_1(b^2_{1}-1) + \lambda_2(b^2_{2}-1)
        \end{aligned}
\end{equation}
The first term in Eq. (47) represents mean field energy for the non-interacting Anderson impurity model with renormalized parameters $(\tilde{h}_{1k,\alpha},\tilde{h}_{2k,\alpha},\tilde{\epsilon}_1,\tilde{\epsilon}_2,\Tilde{t})$. The second and third terms result from the constraint relationship defined by Eq. (10).
Now, to obtain the parameters $\lambda_1$, $\lambda_2$, $b_1$, and $b_2$ we have to minimize the above equation which is done as follows.
\begin{equation}
    \begin{aligned}
        \frac{\partial F_{MF}}{\partial \lambda_1} = & -\frac{N}{\pi}\int_{-D}^{D} \frac{f(\omega)}{(\tilde{\epsilon}_{d1}-\omega^2)+\tilde{\Lambda}^2}\left[ (\tilde{\epsilon}_{d1}-\omega) \frac{\partial \Tilde{\Lambda}}{\partial \lambda_1}
        -\tilde{\Lambda}(\tilde{\epsilon}_{d1}-\omega)\right]
        \\ & 
         d{\omega} + (b_1^2-1)=0  
    \end{aligned}
\end{equation}
\begin{equation}
    \begin{aligned}
        \frac{\partial F_{MF}}{\partial b_1} = & -\frac{N}{\pi} \int_{-D}^{D} \left[\frac{f(\omega)}{(\tilde{\epsilon}_{d1}-\omega^2)+\tilde{\Lambda}^2} (\tilde{\epsilon}_{d1}-\omega)\frac{\partial \Tilde{\Lambda}}{\partial b_1} \right] 
        \\ & 
        d{\omega} +2{\lambda_1} b_{1}=0
 \end{aligned}
\end{equation}
\begin{equation}
    \begin{aligned}
        \frac{\partial F_{MF}}{\partial \lambda_2} = & -\frac{N}{\pi}\int_{-D}^{D} \frac{f(\omega)}{(\tilde{\epsilon}_{d1}-\omega^2)+\tilde{\Lambda}^2}\left[ (\tilde{\epsilon}_{d1}-\omega) \frac{\partial \Tilde{\Lambda}}{\partial \lambda_2}\right] 
        \\ & 
        d{\omega} + (b_2^2-1)=0  
    \end{aligned}
\end{equation}
\begin{equation}
    \begin{aligned}
        \frac{\partial F_{MF}}{\partial b_2} = &  -\frac{N}{\pi} \int_{-D}^{D} \frac{f(\omega)}{(\tilde{\epsilon}_{d1}-\omega^2)+\tilde{\Lambda}^2}\left[ (\tilde{\epsilon}_{d1}-\omega) \frac{\partial \Tilde{\Lambda}}{\partial b_2}\right]
        \\ & 
        d{\omega} + 2{\lambda_2} b_{2}=0 
    \end{aligned}
\end{equation}

Thus, from the above-coupled equations (Eq.48-Eq.51), $\lambda_1$, $\lambda_2$, $b_1$, and $b_2$ can be calculated and finally the renormalized parameters $(\tilde{h}_{1k,\alpha},\tilde{h}_{2k,\alpha},\tilde{\epsilon}_1,\tilde{\epsilon}_2,\Tilde{t})$ can be obtained for given input variables.

In the next section, we go over the numerical analysis of the Andreev bound states and Josephson current across the T-shaped and series-configured double quantum dots system for different parameters.
\end{multicols}

\begin{multicols}{2}
[
\begin{center}
    \textbf{III. RESULTS and DISCUSSION}
\end{center}
]
In this section, we provide a numerical study of ABSs and Josephson current, across the double quantum dots Josephson junction in T-shaped and series configurations. We discussed the behavior of ABS and Josephson current on varying the interdot tunneling and dots energy level at infinite-U limit. By equating the denominator of Eq. (21 and 38) (poles of Green's function) to zero, one can analyze the energy of subgap states ABS in T-shaped and series-configured double quantum dots Josephson junction respectively.
\begin{equation}
    \omega-\tilde{\epsilon}_{d1}-\Tilde{\Lambda}=0
\end{equation}
In Superconductor-quantum dot systems, Josephson current is carried by these subgap states. The Josephson current is calculated by differentiating the ABS with respect to the superconducting phase difference $\phi$, i.e.
 \begin{equation}
     I_{SC}=\frac{\partial E_{ABS}}{\partial \phi}
\end{equation}
We have taken all the parameters in the unit of $\Gamma_{\alpha}$. For simplicity, we assume both the superconducting leads are identical i.e. $\Delta_L=\Delta_R=\Delta$ and also considered symmetric tunneling between $QD's$ and superconducting leads i.e. $\Gamma_{L}=\Gamma_{R}=\Gamma$.
\end{multicols}
\begin{center}
    \textbf{A. T shape configuration}
\end{center}

\begin{figure}[!ht]
  \begin{center}
    \includegraphics[width=0.8\textwidth]{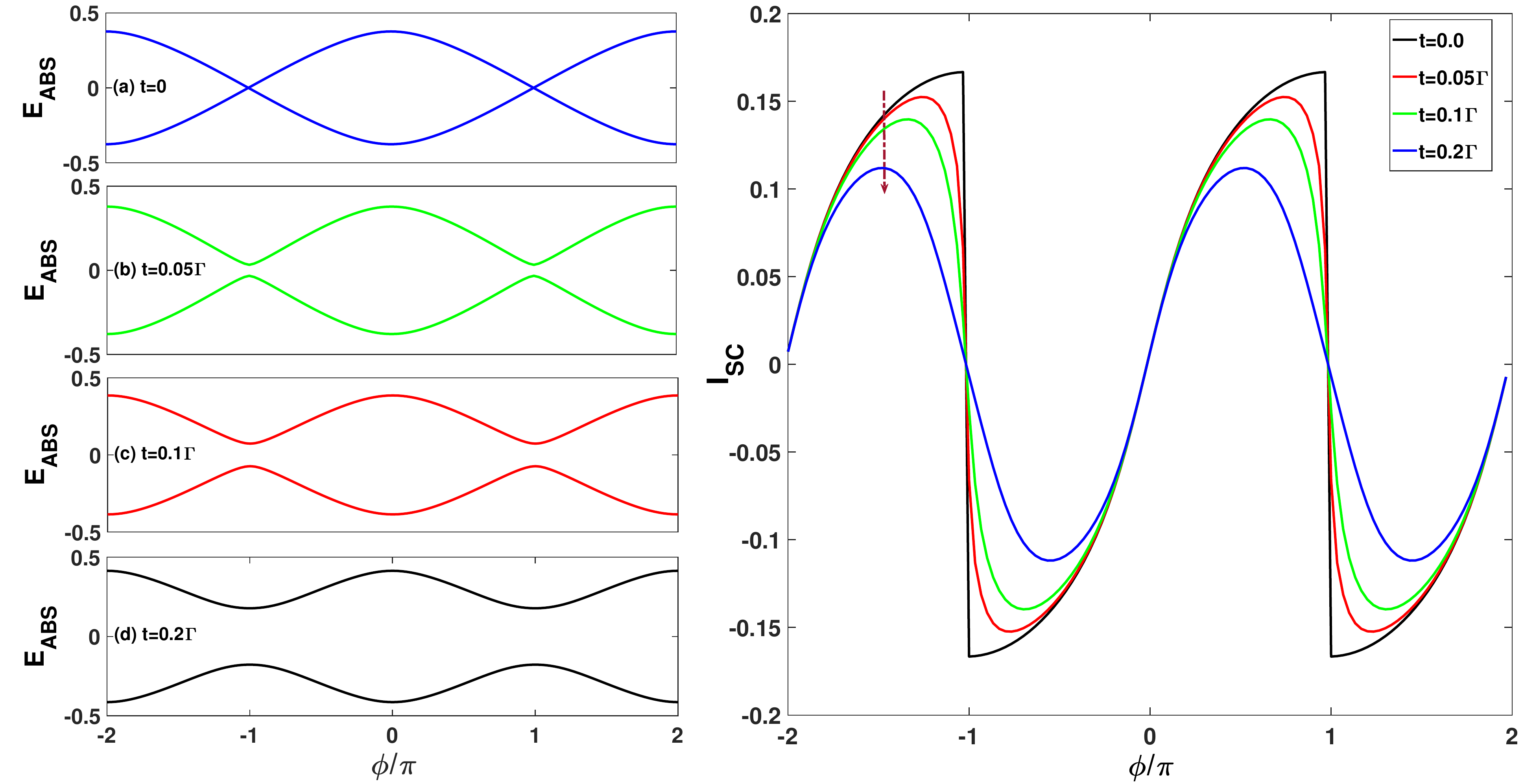}
  \end{center}
 \caption{\label{fig:Tshape_1} \textbf{left:} Energy of Andreev Bound states ($E_{ABS}$) and \textbf{right:} Josephson current $I_{SC}$ ($e/h$) versus superconducting phase difference ($\phi$) for various values of interdot tunneling. The other parameters are $\Delta =0.5\Gamma$, $\epsilon_{d1}=\epsilon_{d2}=0$}
\end{figure}

\begin{figure}[!hbt]
  \begin{center}
    \includegraphics[width=0.8\textwidth]{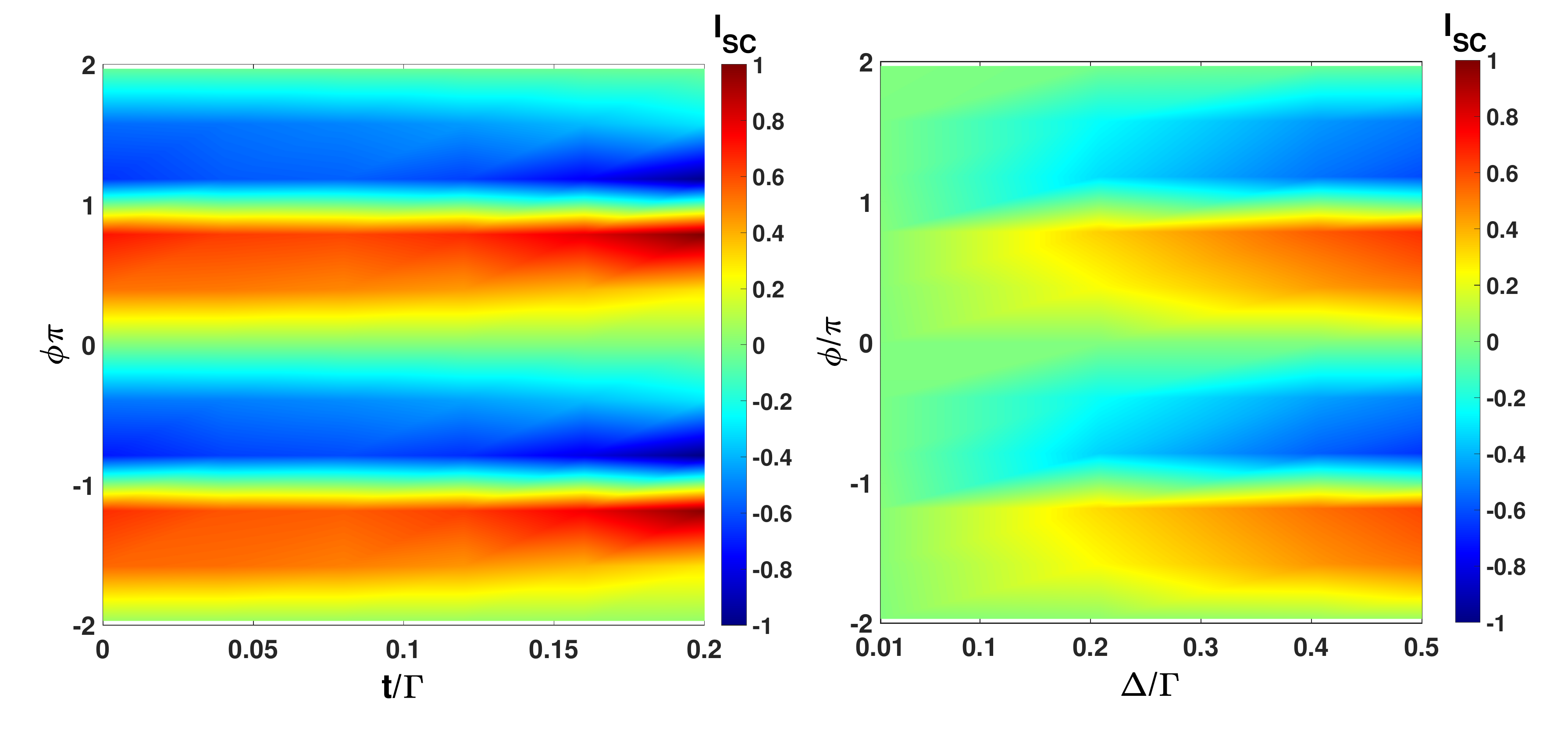}
  \end{center}
 \caption{\label{fig:Tshape_2} 2D phase diagrams of Josephson current as a function of interdot tunneling and $\phi$ at $\Delta/\Gamma=0.5$ (\textbf{left}), and superconducting gap and $\phi$ at $t/\Gamma=0.2$ (\textbf{right})}
\end{figure}
\begin{multicols}{2}  
In figure \ref{fig:Tshape_1}, a plot of energy of Andreev bound states (left) and Josephson current (right) versus superconducting phase difference is shown for various values of the interdot tunneling parameter (t). For $t=0$, i.e. $QD_2$ is not coupled with $QD_1$, the systems show the nature of the S-QD-S system. In S-QD-S systems cooper pairs transport from one lead to another lead due to the Andreev reflection mechanism at the dot-lead interface. Here, lower and upper ABS crosses Fermi energy $E_F=0$ and Josephson current exhibits a discontinuity at $\phi=\pm \pi$. For $t=0$, systems behave as a perfect transmitting channel. Thus, For $t>0$, i.e. $QD_2$ is coupled with $QD_1$, ABS departs from Fermi energy $E_F=0$, and Josephson current suppresses and shows sinusoidal nature. For coupled quantum dots, as t increases, the equivalent energy level ($\epsilon_{di} \pm t$) of dots splits into two, each moving away from the Fermi energy and supercurrent suppresses. Therefore, the system does not function as a perfect transmitting channel in the case of coupled quantum dots. The explanation of suppression of Josephson current may also be stated as follows. For coupled quantum dots electrons tend to tunnel into $QD_2$ with increasing t. Because of this, interference destruction occurs between two transport paths, and Josephson current suppresses with increasing t.
Figure \ref{fig:Tshape_2}, presents the 2D phase diagrams of Josephson current as a function of  $t/\Gamma, \phi$ (left) and $\Delta/\Gamma, \phi$ (right). 

\begin{figure}[!ht]
  \begin{center}
    \includegraphics[width=0.8\textwidth]{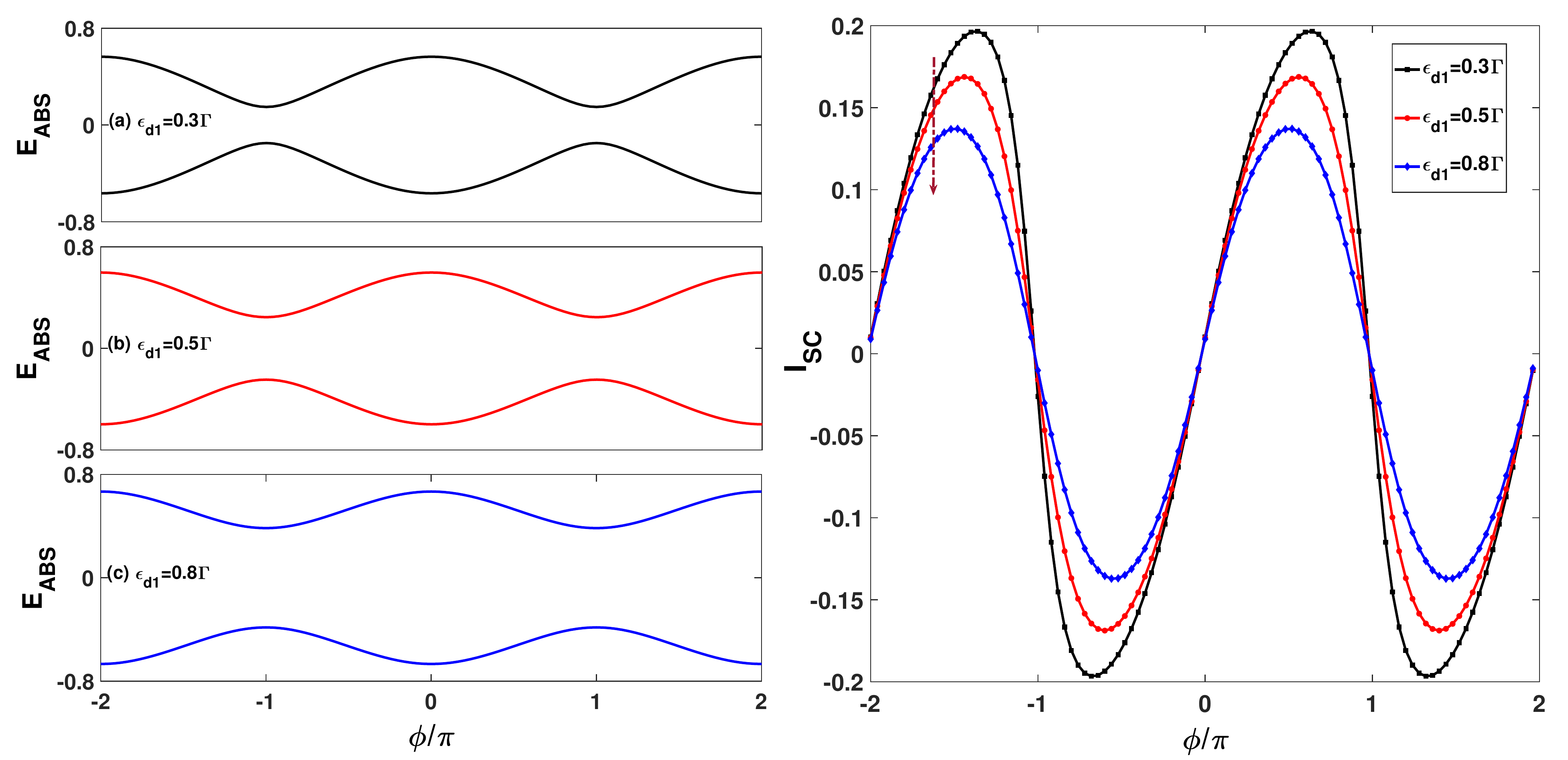}
  \end{center}
 \caption{\label{fig:Tshape_3} \textbf{left:}  Energy of Andreev Bound states ($E_{ABS}$) versus superconducting phase difference ($\phi$) for different values of $QD_1$ energy levels, \textbf{right}: corresponding Josephson current $I_{SC}$ ($e/h$). Other parameters are $\Delta =0.5\Gamma$,  $\epsilon_{d2}=0$, $t=0.02\Gamma$.}
\end{figure}

 In figure \ref{fig:Tshape_3}, a plot of Andreev bound states (left) and Josephson current (right) versus superconducting phase difference is shown for various values of $QD_1$ energy levels. There is a finite gap between lower and upper ABS when energy levels of $QD_1$ lie above the Fermi level and Josephson current suppress with a sinusoidal nature. As the main quantum dot energy levels lie above the Fermi level, it acts as a trap for the quasiparticles involved in Andreev reflection at the dot-lead interface. Due to this, the spectral weight of ABS reduces leading to a decrease in Josephson current. The reduction of spectral weight refers that fewer quasiparticles can tunnel into the ABSs and carry the Josephson current.

\begin{figure}[!ht]
  \begin{center}
    \includegraphics[width=0.8\textwidth]{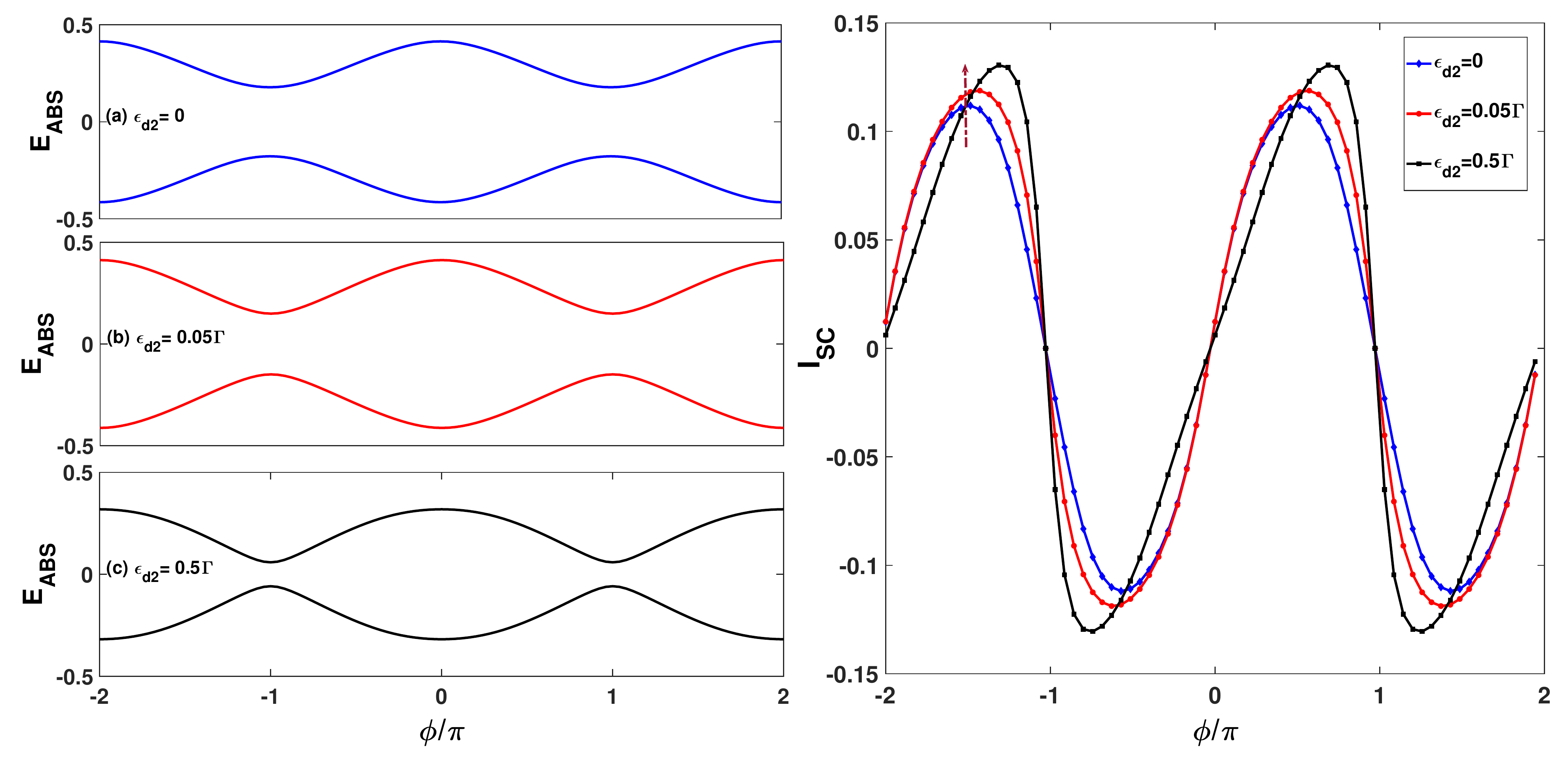}
  \end{center}
 \caption{\label{fig:Tshape_4} \textbf{left:} Energy of Andreev Bound states ($E_{ABS}$)  and \textbf{right}: Josephson current $I_{SC}$ ($e/h$) versus superconducting phase difference ($\phi$) for different values of $QD_2$ energy levels. Other parameters are $\Delta =0.5\Gamma$,  $\epsilon_{d1}=0$, $t=0.2\Gamma$}
\end{figure}

In figure \ref{fig:Tshape_4}, we plot ABS (left) and Josephson current (right) versus the superconducting phase difference for different $QD_2$ energy levels. The Josephson current is enhanced as the $QD_2$ energy levels lie above the Fermi level and suppress at the Fermi level $\epsilon_{d2}=0$. This enhancement in Josephson current is due to the additional transport channel provided by the side dot, which increases the spectral weight of the ABS, resulting in a larger Josephson current. The gap between lower and upper ABS decreases as $\epsilon_{d2}$ departs from the Fermi level.
\end{multicols}

 \begin{center}
    \textbf{B. Series configuration}
\end{center}

\begin{figure}[!ht]
  \begin{center}
    \includegraphics[width=0.8\textwidth]{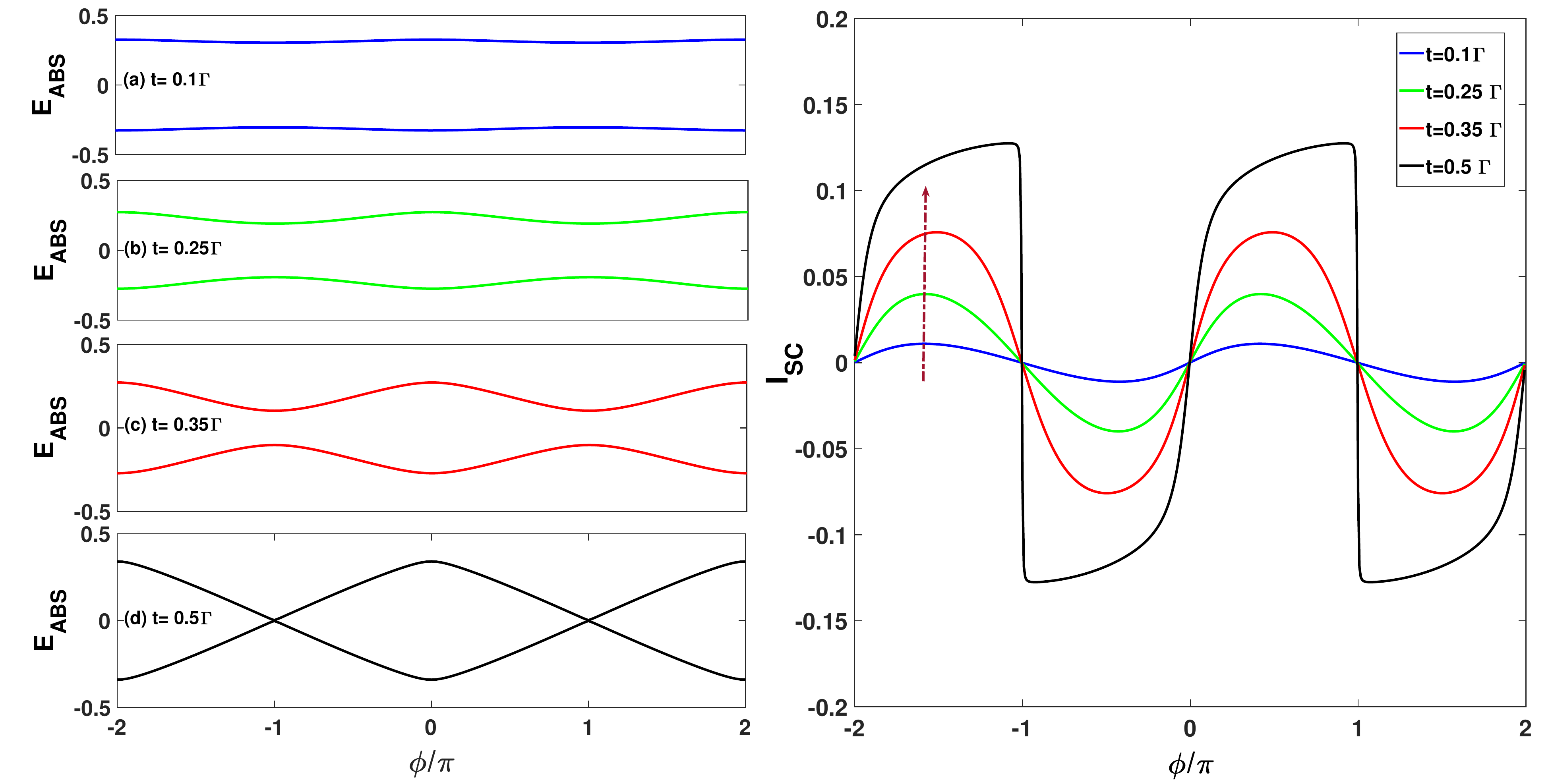}
  \end{center}
 \caption{\label{fig:series_1} \textbf{left:} Energy of Andreev Bound states ($E_{ABS}$) and \textbf{right:} Josephson current $I_{SC}$ ($e/h$) versus superconducting phase difference ($\phi$) for various values of interdot tunneling. The other parameters are $\Delta =0.5\Gamma$,  $\epsilon_{d1}=\epsilon_{d2}=0.5\Gamma$}
\end{figure}
\begin{figure}[!hbt]
  \begin{center}
    \includegraphics[width=0.8\textwidth]{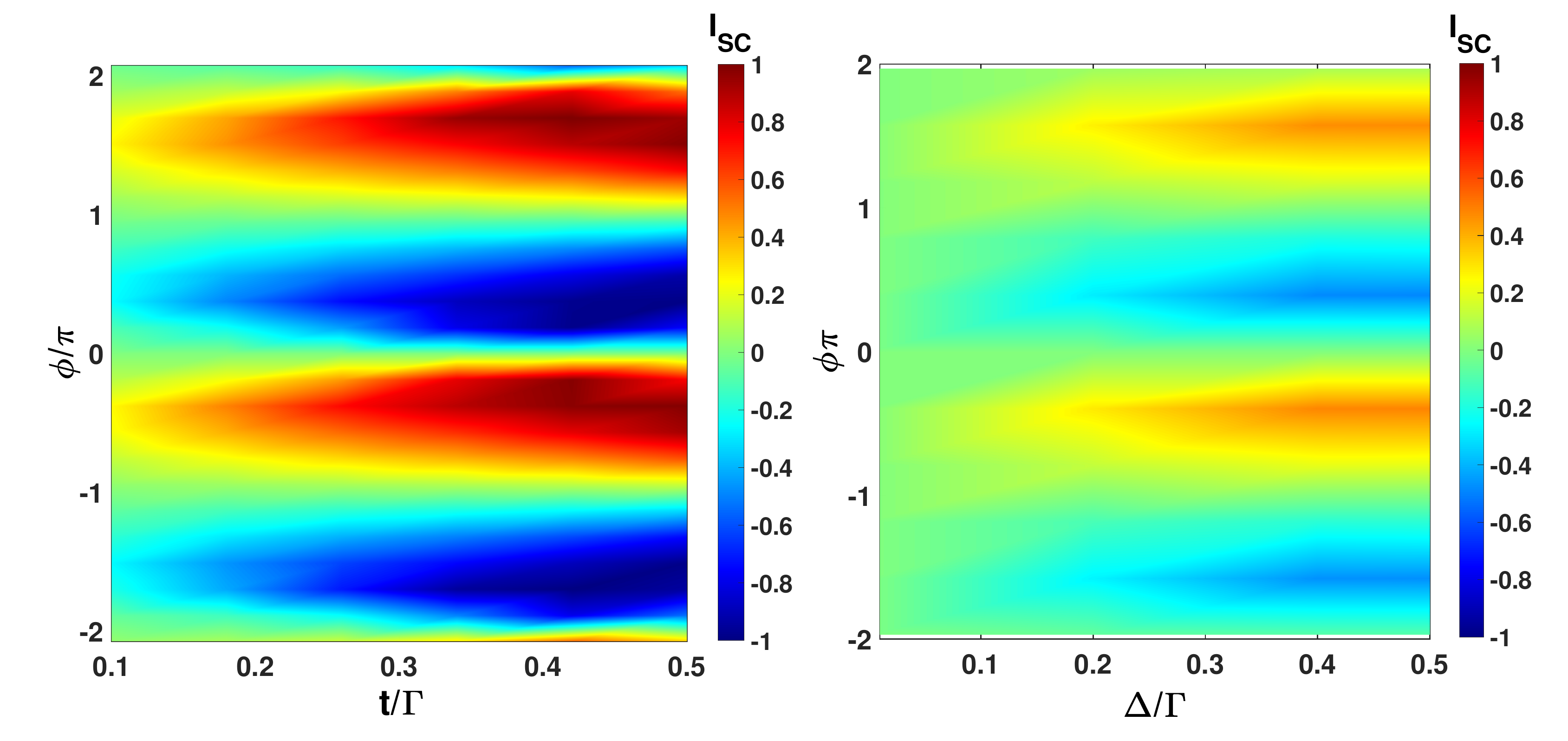}
  \end{center}
 \caption{\label{fig:series_2} \textbf{left:} 2D phase diagrams of Josephson current as a function of interdot tunneling and $\phi$ at $\Delta/\Gamma=0.5$ (\textbf{left}), and superconducting gap and $\phi$ at $t/\Gamma=0.35$ (\textbf{right})}
\end{figure}
\begin{figure}[!t]
  \begin{center}
    \includegraphics[width=0.8\textwidth]{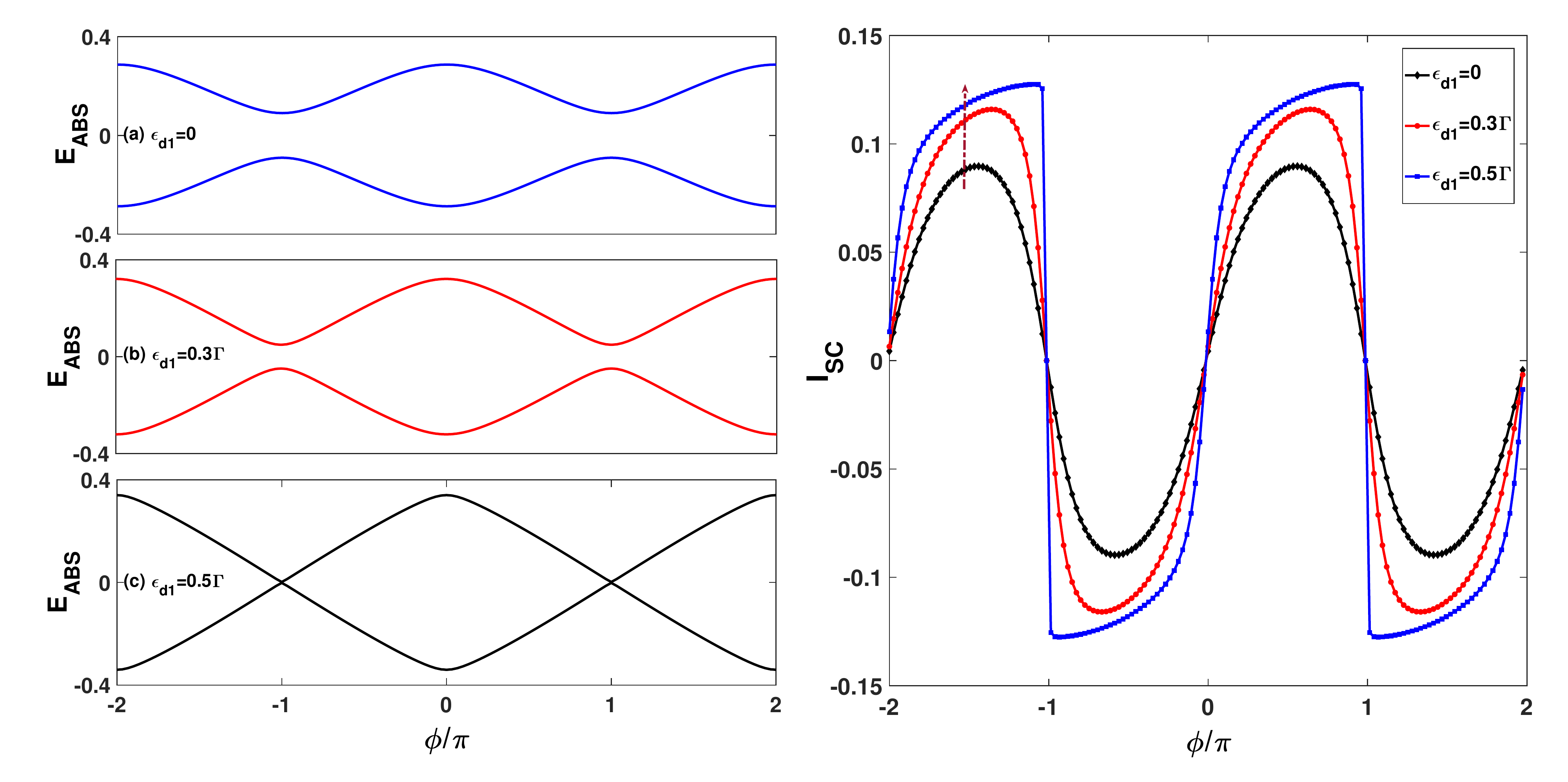}
  \end{center}
 \caption{\label{fig:series_3} \textbf{left:} Energy of Andreev Bound states ($E_{ABS}$) versus superconducting phase difference ($\phi$) for different values of $QD_1$ energy levels, \textbf{right}: corresponding Josephson current $I_{SC}$ ($e/h$). Other parameters are $\Delta =0.5\Gamma$, $\epsilon_{d2}=0.5$, $t=0.5\Gamma$.}
\end{figure}
\begin{figure}[!hbt]
  \begin{center}
    \includegraphics[width=0.8\textwidth]{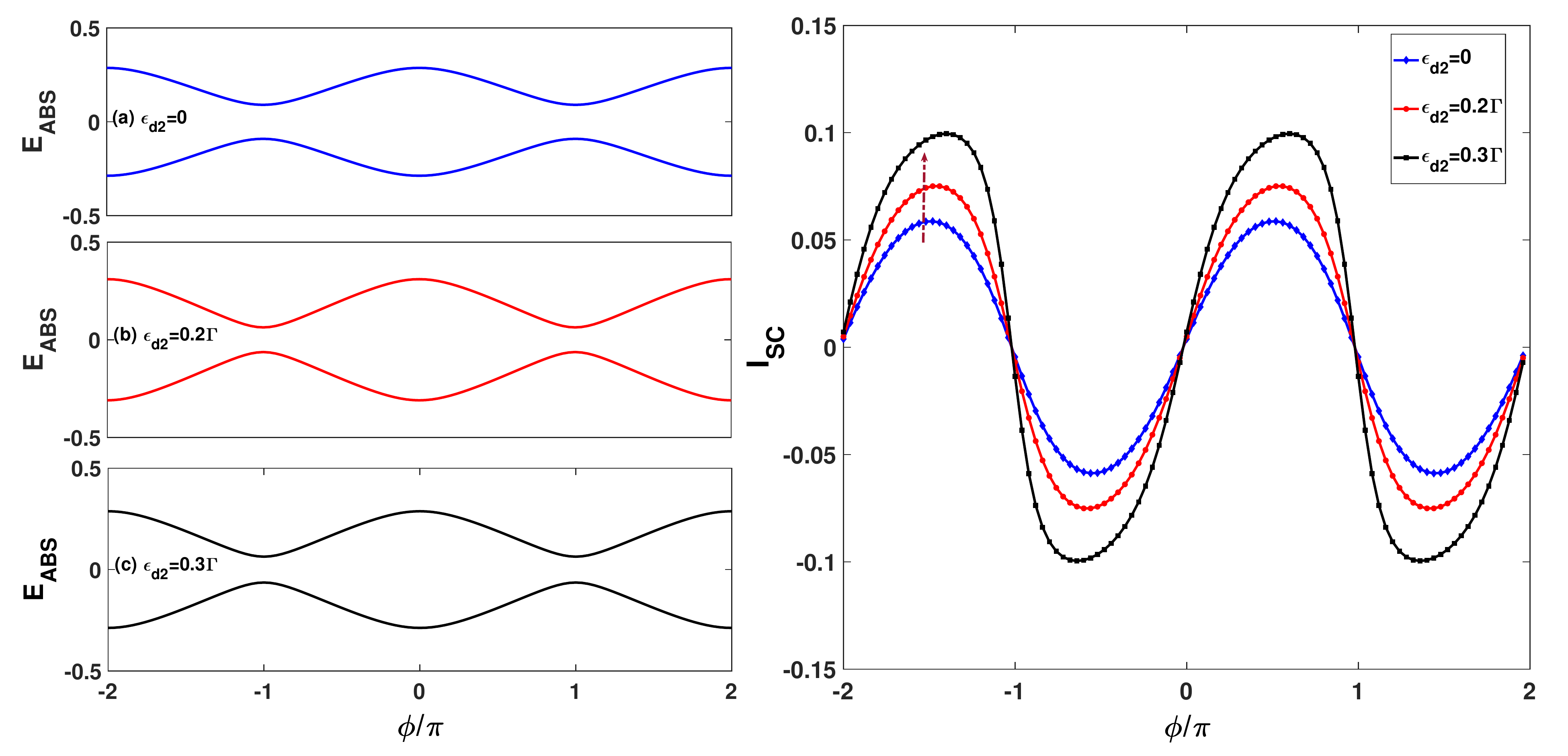}
  \end{center}
 \caption{\label{fig:series_4} \textbf{left:} Energy of Andreev Bound states ($E_{ABS}$)  and \textbf{right}: Josephson current $I_{SC}$ ($e/h$) versus superconducting phase difference ($\phi$) for different values of $QD_2$ energy levels. Other parameters are $\Delta =0.5\Gamma$, $\epsilon_{d1}=0.5$, $t=0.5\Gamma$}
\end{figure}

\begin{multicols}{2}
In figure \ref{fig:series_1}, a plot of Andreev Bound states (left) and Josephson current (right) is shown as a function of superconducting phase difference for different values of interdot tunneling. For decoupled quantum dots (t=0), there is no sign of ABS in the series configuration. For coupled quantum dots $t>0$, the lower and upper ABS exhibits finite gap from Fermi energy and at $t=0.5\Gamma$ ABS cross at Fermi energy $\omega=0$. Further increment in $t$ leads to a finite gap in lower and upper ABS. The supercurrent increases with increasing t and shows sinusoidal nature. For $t=0.5\Gamma$, there is resonance tunneling between dot's energy level and lead, and the Josephson current shows a discontinuity at $\phi=\pm \pi$. The results of the series configuration are opposite from the T-shape and parallel configuration where Josephson current decreases with increasing t.   
Figure \ref{fig:series_2}, presents the 2D phase diagrams of Josephson current as a function of  $t/\Gamma, \phi$ (left) and $\Delta/\Gamma, \phi$ (right).

In figure \ref{fig:series_3}, a plot of Andreev bound states (left) and Josephson current (right) versus superconducting phase difference is shown for various values of $QD_1$ energy levels. There is a finite gap between lower and upper ABS when energy levels of $QD_1$ lie above the Fermi level and cross $\omega=0$ for $\epsilon_{d1}=0.5$. The Josephson current increases with increasing energy levels of $QD_1$ and maximum for $\epsilon_{d1}=\epsilon_{d2}=0.5$ and $t=0.5$. With the increasing $QD_1$ energy levels the tunneling probability of electrons through the junction increases and thus the Josephson current increases. For the $\epsilon_{d1}=\epsilon_{d1}=t=0.5\Gamma$ case, the resonant tunneling occurs between dots and lead, and the Josephson current is maximum.

In figure \ref{fig:series_4}, we plot ABS (left) and Josephson current (right) versus the superconducting phase difference for different energy levels of $QD_2$. At the Fermi level of $QD_2$ ($\epsilon_{d2}=0$), the Josephson current is suppressed, while it is enhanced as $QD_2$ energy levels lie above the Fermi level. The reason for the enhancement of the Josephson current is already discussed in the previous paragraph. The gap between lower and upper ABS decreases as $\epsilon_{d2}$ departs from the Fermi level.
\end{multicols}

\begin{center}
    \textbf{IV. CONCLUSION}
\end{center}
\begin{multicols}{2}
We have addressed the Josephson transport across the double quantum dot Josephson junction in T-shape and series configuration. It is observed that interdot tunneling and energy levels of quantum dots perform a key role in the tunability of ABSs, and Josephson current. The Josephson current decreases with increasing interdot tunneling in a T-shaped configuration while enhancing in a series configuration with increasing interdot tunneling. The subgap ABS also exhibits the opposite behavior in both configurations. In T-shaped configuration, subgap ABS cross at Fermi energy $\omega=0$ for decoupled quantum dots and shows sinusoidal nature for coupled quantum dots. In series configuration, subgap ABS crosses Fermi energy when there is resonant tunneling between the dot's energy level i.e. $t=0.5\Gamma$ and $\epsilon_{d1}=\epsilon_{d2}=0.5$. We have also analyzed the nature of the energy of ABS and Josephson current with the energy levels of quantum dots for the T-shape and series configuration. As $QD_1$ and $QD_2$ energy levels rise above the Fermi level, the subgap ABS and Josephson current show different behaviors in the T-shape configuration. The Josephson current decreases as the $QD_1$ energy level rises above the Fermi level, while it increases as the $QD_2$ energy level rises above Fermi's level. In the series configuration, the Josephson current increases as $QD_1$ and $QD_2$ energy levels lie above the Fermi level.

The study of Andreev bound states and Josephson current can be useful in quantum computing and nanoelectronics. Controlling the Andreev bound states and Josephson current in double quantum dot-based Josephson junctions may improve the efficiency and robustness of quantum processing \cite{wendin2007quantum}. Josephson junctions are key building blocks of superconducting electronics, which have extensive uses in sensing, metrology, and communication. The double quantum dot-based Josephson junction could offer tunable supercurrents and novel noise properties in future superconducting devices \cite{arnault2022dynamical,PhysRevB.98.035438,PhysRevResearch.3.033240,PRXQuantum.3.030311}. This study can also be extended for multi-dot and multi-terminal Josephson junctions.
\end{multicols}

\begin{multicols}{2}
\textbf{Acknowledgement}
Authors offer their gratitude to the DST-SER-1644-PHY 2021-22 research project for financial support. Bhupendra Kumar, a research scholar in the Department of Physics, IIT Roorkee, also acknowledges the support from the Ministry of Education (MoE), India in the form of a Ph.D. fellowship.
\end{multicols}

\bibliographystyle{elsarticle-num} 
\bibliography{Mybib3.bib}

\begin{thebibliography}{10}
\expandafter\ifx\csname url\endcsname\relax
  \def\url#1{\texttt{#1}}\fi
\expandafter\ifx\csname urlprefix\endcsname\relax\def\urlprefix{URL }\fi
\expandafter\ifx\csname href\endcsname\relax
  \def\href#1#2{#2} \def\path#1{#1}\fi

\bibitem{Josephson1962}
B.~D. Josephson, Possible new effects in superconductive tunnelling, Physics
  Letters 1 (1962) 251--253.
\newblock \href {https://doi.org/10.1016/0031-9163(62)91369-0}
  {\path{doi:10.1016/0031-9163(62)91369-0}}.

\bibitem{anderson1970josephson}
P.~W. Anderson, How josephson discovered his effect, Phys. Today 23~(11) (1970)
  23--29.
\newblock \href {https://doi.org/10.1063/1.3021826}
  {\path{doi:10.1063/1.3021826}}.

\bibitem{kouwenhoven2001few}
Kouwenhoven, et~al., Few-electron quantum dots, Reports on Progress in Physics
  64~(6) (2001) 701.
\newblock \href {https://doi.org/10.1088/0034-4885/64/6/201}
  {\path{doi:10.1088/0034-4885/64/6/201}}.

\bibitem{Kastner1993}
M.~A. Kastner, Artificial atoms, Physics Today 46 (1993).
\newblock \href {https://doi.org/10.1063/1.881393}
  {\path{doi:10.1063/1.881393}}.

\bibitem{martin2011josephson}
A.~Mart{\'\i}n-Rodero, A.~Levy~Yeyati, Josephson and andreev transport through
  quantum dots, Advances in Physics 60~(6) (2011) 899--958.
\newblock \href {https://doi.org/10.1038/nnano.2010.173}
  {\path{doi:10.1038/nnano.2010.173}}.

\bibitem{de2010hybrid}
D.~Franceschi, et~al., Hybrid superconductor--quantum dot devices, Nature
  nanotechnology 5~(10) (2010) 703--711.
\newblock \href {https://doi.org/10.1038/nnano.2010.173}
  {\path{doi:10.1038/nnano.2010.173}}.

\bibitem{PhysRevLett.99.126602}
Eichler, et~al., Even-odd effect in andreev transport through a carbon nanotube
  quantum dot, Phys. Rev. Lett. 99 (2007) 126602.
\newblock \href {https://doi.org/10.1103/PhysRevLett.99.126602}
  {\path{doi:10.1103/PhysRevLett.99.126602}}.

\bibitem{PhysRevLett.89.256801}
M.~R. Buitelaar, et~al., Quantum dot in the kondo regime coupled to
  superconductors, Phys. Rev. Lett. 89 (2002) 256801.
\newblock \href {https://doi.org/10.1103/PhysRevLett.89.256801}
  {\path{doi:10.1103/PhysRevLett.89.256801}}.

\bibitem{sand2007kondo}
T.~Sand-Jespersen, et~al., Kondo-enhanced andreev tunneling in inas nanowire
  quantum dots, Physical review letters 99~(12) (2007) 126603.
\newblock \href {https://doi.org/10.1103/PhysRevLett.99.126603}
  {\path{doi:10.1103/PhysRevLett.99.126603}}.

\bibitem{r10}
K.~G. Rasmussen, et~al., Kondo resonance enhanced supercurrent in single wall
  carbon nanotube josephson junctions, New Journal of Physics 9 (2007).
\newblock \href {https://doi.org/10.1088/1367-2630/9/5/124}
  {\path{doi:10.1088/1367-2630/9/5/124}}.

\bibitem{Cleuziou2006}
J.~P. Cleuziou, et~al., Carbon nanotube superconducting quantum interference
  device, Nature Nanotechnology 1 (2006).
\newblock \href {https://doi.org/10.1038/nnano.2006.54}
  {\path{doi:10.1038/nnano.2006.54}}.

\bibitem{PhysRevX.2.011009}
R.~Maurand, et~al., First-order $0-\pi$ quantum phase transition in the kondo
  regime of a superconducting carbon-nanotube quantum dot, Phys. Rev. X 2
  (2012) 011009.
\newblock \href {https://doi.org/10.1103/PhysRevX.2.011009}
  {\path{doi:10.1103/PhysRevX.2.011009}}.

\bibitem{r13}
J.~A.~V. Dam, et~al., Supercurrent reversal in quantum dots, Nature 442 (2006).
\newblock \href {https://doi.org/10.1038/nature05018}
  {\path{doi:10.1038/nature05018}}.

\bibitem{Verma2020}
S.~Verma, Ajay, Influence of superconductivity on the magnetic moment of
  quantum impurity embedded in bcs superconductor, Journal of Physics Condensed
  Matter 33 (2020).
\newblock \href {https://doi.org/10.1088/1361-648X/abcc0e}
  {\path{doi:10.1088/1361-648X/abcc0e}}.

\bibitem{PhysRevLett.91.057005}
Buitelaar, et~al., Multiple andreev reflections in a carbon nanotube quantum
  dot, Phys. Rev. Lett. 91 (2003) 057005.
\newblock \href {https://doi.org/10.1103/PhysRevLett.91.057005}
  {\path{doi:10.1103/PhysRevLett.91.057005}}.

\bibitem{r15}
P.~Jarillo-Herrero, et~al., Quantum supercurrent transistors in carbon
  nanotubes, Nature 439 (2006).
\newblock \href {https://doi.org/10.1038/nature04550}
  {\path{doi:10.1038/nature04550}}.

\bibitem{PhysRevLett.96.207003}
H.~I. J\o{}rgensen, et~al., Electron transport in single-wall carbon nanotube
  weak links in the fabry-perot regime, Phys. Rev. Lett. 96 (2006) 207003.
\newblock \href {https://doi.org/10.1103/PhysRevLett.96.207003}
  {\path{doi:10.1103/PhysRevLett.96.207003}}.

\bibitem{doi:10.1021/nl071152w}
J.~H. Ingerslev, et~al., Critical current $0−\pi$ transition in designed
  josephson quantum dot junctions, Nano Letters 7~(8) (2007) 2441--2445.
\newblock \href {https://doi.org/10.1021/nl071152w}
  {\path{doi:10.1021/nl071152w}}.

\bibitem{PhysRevB.79.155441}
H.~I. J\o{}rgensen, et~al., Critical and excess current through an open quantum
  dot: Temperature and magnetic-field dependence, Phys. Rev. B 79 (2009)
  155441.
\newblock \href {https://doi.org/10.1103/PhysRevB.79.155441}
  {\path{doi:10.1103/PhysRevB.79.155441}}.

\bibitem{Pillet2010}
J.~D. Pillet, et~al., Andreev bound states in supercurrent-carrying carbon
  nanotubes revealed, Nature Physics 6 (2010).
\newblock \href {https://doi.org/10.1038/nphys1811}
  {\path{doi:10.1038/nphys1811}}.

\bibitem{Hofstetter2009}
L.~Hofstetter, et~al., Cooper pair splitter realized in a two-quantum-dot
  y-junction, Nature 461 (2009).
\newblock \href {https://doi.org/10.1038/nature08432}
  {\path{doi:10.1038/nature08432}}.

\bibitem{PhysRevLett.104.026801}
L.~G. Herrmann, et~al., Carbon nanotubes as cooper-pair beam splitters, Phys.
  Rev. Lett. 104 (2010) 026801.
\newblock \href {https://doi.org/10.1103/PhysRevLett.104.026801}
  {\path{doi:10.1103/PhysRevLett.104.026801}}.

\bibitem{vecino2003josephson}
E.~Vecino, et~al., Josephson current through a correlated quantum level:
  Andreev states and $\pi$ junction behavior, Physical Review B 68~(3) (2003)
  035105.
\newblock \href {https://doi.org/10.1103/PhysRevB.68.035105}
  {\path{doi:10.1103/PhysRevB.68.035105}}.

\bibitem{choi2004kondo}
M.-S. Choi, et~al., Kondo effect and josephson current through a quantum dot
  between two superconductors, Physical Review B 70~(2) (2004) 020502.
\newblock \href {https://doi.org/10.1103/PhysRevB.70.020502}
  {\path{doi:10.1103/PhysRevB.70.020502}}.

\bibitem{lim2008andreev}
J.~S. Lim, et~al., Andreev bound states in the kondo quantum dots coupled to
  superconducting leads, Journal of Physics: Condensed Matter 20~(41) (2008)
  415225.
\newblock \href {https://doi.org/10.1088/0953-8984/20/41/415225}
  {\path{doi:10.1088/0953-8984/20/41/415225}}.

\bibitem{zhu2001andreev}
Y.~Zhu, et~al., Andreev bound states and the $\pi$-junction transition in a
  superconductor/quantum-dot/superconductor system, Journal of Physics:
  Condensed Matter 13~(39) (2001) 8783.
\newblock \href {https://doi.org/10.1088/0953-8984/13/39/307}
  {\path{doi:10.1088/0953-8984/13/39/307}}.

\bibitem{karrasch2008josephson}
K.~Christoph, et~al., Josephson current through a single anderson impurity
  coupled to bcs leads, Physical Review B 77~(2) (2008) 024517.
\newblock \href {https://doi.org/10.1103/PhysRevB.77.024517}
  {\path{doi:10.1103/PhysRevB.77.024517}}.

\bibitem{Cheng_2008}
S.~guang Cheng, Q.~feng Sun, Josephson current transport through t-shaped
  double quantum dots, Journal of Physics: Condensed Matter 20~(50) (2008)
  505202.
\newblock \href {https://doi.org/10.1088/0953-8984/20/50/505202}
  {\path{doi:10.1088/0953-8984/20/50/505202}}.

\bibitem{PhysRevB.75.045132}
L.~Rosa, et~al., Josephson current through a kondo molecule, Phys. Rev. B 75
  (2007) 045132.
\newblock \href {https://doi.org/10.1103/PhysRevB.75.045132}
  {\path{doi:10.1103/PhysRevB.75.045132}}.

\bibitem{PhysRevB.107.115165}
Ortega-Taberner, et~al., Anomalous josephson current through a driven double
  quantum dot, Phys. Rev. B 107 (2023) 115165.
\newblock \href {https://doi.org/10.1103/PhysRevB.107.115165}
  {\path{doi:10.1103/PhysRevB.107.115165}}.

\bibitem{Chi2005}
F.~Chi, S.~S. Li, Current-voltage characteristics in strongly correlated double
  quantum dots, Journal of Applied Physics 97 (2005).
\newblock \href {https://doi.org/10.1063/1.1939065}
  {\path{doi:10.1063/1.1939065}}.

\bibitem{PhysRevB.66.085306}
Z.~Yu, et~al., Probing spin states of coupled quantum dots by a dc josephson
  current, Phys. Rev. B 66 (2002) 085306.
\newblock \href {https://doi.org/10.1103/PhysRevB.66.085306}
  {\path{doi:10.1103/PhysRevB.66.085306}}.

\bibitem{Droste2012}
S.~Droste, et~al., Josephson current through interacting double quantum dots
  with spin-orbit coupling, Journal of Physics Condensed Matter 24 (2012).
\newblock \href {https://doi.org/10.1088/0953-8984/24/41/415301}
  {\path{doi:10.1088/0953-8984/24/41/415301}}.

\bibitem{RAJPUT2014193}
G.~Rajput, R.~Kumar, Ajay, Tunable josephson effect in hybrid parallel coupled
  double quantum dot-superconductor tunnel junction, Superlattices and
  Microstructures 73 (2014) 193--202.
\newblock \href {https://doi.org/https://doi.org/10.1016/j.spmi.2014.05.029}
  {\path{doi:https://doi.org/10.1016/j.spmi.2014.05.029}}.

\bibitem{PhysRevLett.105.116803}
R.~Žitko, et~al., Josephson current in strongly correlated double quantum
  dots, Phys. Rev. Lett. 105 (2010) 116803.
\newblock \href {https://doi.org/10.1103/PhysRevLett.105.116803}
  {\path{doi:10.1103/PhysRevLett.105.116803}}.

\bibitem{r34}
J.~C.~E. Saldaña, et~al., Supercurrent in a double quantum dot, Physical
  Review Letters 121 (2018).
\newblock \href {https://doi.org/10.1103/PhysRevLett.121.257701}
  {\path{doi:10.1103/PhysRevLett.121.257701}}.

\bibitem{haug2008quantum}
H.~Haug, A.-P. Jauho, et~al., Quantum kinetics in transport and optics of
  semiconductors, Vol.~2, Springer, Berlin, 2008.
\newblock \href {https://doi.org/10.1007/978-3-540-73564-9}
  {\path{doi:10.1007/978-3-540-73564-9}}.

\bibitem{keldysh1965diagram}
L.~V. Keldysh, et~al., Diagram technique for nonequilibrium processes, Sov.
  Phys. JETP 20~(4) (1965) 1018--1026.

\bibitem{PhysRevLett.108.227001}
D.~J. Luitz, et~al., Understanding the josephson current through a
  kondo-correlated quantum dot, Phys. Rev. Lett. 108 (2012) 227001.
\newblock \href {https://doi.org/10.1103/PhysRevLett.108.227001}
  {\path{doi:10.1103/PhysRevLett.108.227001}}.

\bibitem{Probst2016}
B.~Probst, et~al., Signatures of nonlocal cooper-pair transport and of a
  singlet-triplet transition in the critical current of a double-quantum-dot
  josephson junction, Physical Review B 94 (2016).
\newblock \href {https://doi.org/10.1103/PhysRevB.94.155445}
  {\path{doi:10.1103/PhysRevB.94.155445}}.

\bibitem{PhysRevB.92.014504}
B.~Sothmann, et~al., Josephson response of a conventional and a
  noncentrosymmetric superconductor coupled via a double quantum dot, Phys.
  Rev. B 92 (2015) 014504.
\newblock \href {https://doi.org/10.1103/PhysRevB.92.014504}
  {\path{doi:10.1103/PhysRevB.92.014504}}.

\bibitem{Wang2017}
X.~Q. Wang, G.~Y. Yi, W.~J. Gong, Effect of interdot coulomb interaction on the
  josephson phase transition in a double-quantum-dot junction, Superlattices
  and Microstructures 109 (2017).
\newblock \href {https://doi.org/10.1016/j.spmi.2017.05.026}
  {\path{doi:10.1016/j.spmi.2017.05.026}}.

\bibitem{PhysRevB.62.13569}
M.-S. Choi, et~al., Spin-dependent josephson current through double quantum
  dots and measurement of entangled electron states, Phys. Rev. B 62 (2000)
  13569--13572.
\newblock \href {https://doi.org/10.1103/PhysRevB.62.13569}
  {\path{doi:10.1103/PhysRevB.62.13569}}.

\bibitem{PhysRevB.89.235110}
J.~F. Rentrop, et~al., Nonequilibrium transport through a josephson quantum
  dot, Phys. Rev. B 89 (2014) 235110.
\newblock \href {https://doi.org/10.1103/PhysRevB.89.235110}
  {\path{doi:10.1103/PhysRevB.89.235110}}.

\bibitem{PhysRevB.95.045104}
T.~Domański, et~al., Josephson-phase-controlled interplay between correlation
  effects and electron pairing in a three-terminal nanostructure, Phys. Rev. B
  95 (2017) 045104.
\newblock \href {https://doi.org/10.1103/PhysRevB.95.045104}
  {\path{doi:10.1103/PhysRevB.95.045104}}.

\bibitem{r51}
K.~Grove-Rasmussen, et~al., Yu-shiba-rusinov screening of spins in double
  quantum dots, Nature Communications 9 (2018).
\newblock \href {https://doi.org/10.1038/s41467-018-04683-x}
  {\path{doi:10.1038/s41467-018-04683-x}}.

\bibitem{Baba2017}
S.~Baba, et~al., Gate tunable parallel double quantum dots in inas
  double-nanowire devices, Applied Physics Letters 111 (2017).
\newblock \href {https://doi.org/10.1063/1.4997646}
  {\path{doi:10.1063/1.4997646}}.

\bibitem{PhysRevB.72.033414}
A.~Y. Kasumov, et~al., Proximity effect in a
  superconductor-metallofullerene-superconductor molecular junction, Phys. Rev.
  B 72 (2005) 033414.
\newblock \href {https://doi.org/10.1103/PhysRevB.72.033414}
  {\path{doi:10.1103/PhysRevB.72.033414}}.

\bibitem{RevModPhys.75.1}
W.~G. van~der Wiel, et~al., Electron transport through double quantum dots,
  Rev. Mod. Phys. 75 (2002) 1--22.
\newblock \href {https://doi.org/10.1103/RevModPhys.75.1}
  {\path{doi:10.1103/RevModPhys.75.1}}.

\bibitem{Nilsson2017}
M.~Nilsson, et~al., Parallel-coupled quantum dots in inas nanowires, Nano
  Letters 17 (2017).
\newblock \href {https://doi.org/10.1021/acs.nanolett.7b04090}
  {\path{doi:10.1021/acs.nanolett.7b04090}}.

\bibitem{r39}
H.~I. Jørgensen, et~al., Critical current $0-\pi$ transition in designed
  josephson quantum dot junctions, Nano Letters 7 (2007).
\newblock \href {https://doi.org/10.1021/nl071152w}
  {\path{doi:10.1021/nl071152w}}.

\bibitem{PhysRevLett.128.046801}
P.~Zhang, et~al., Signatures of andreev blockade in a double quantum dot
  coupled to a superconductor, Phys. Rev. Lett. 128 (2022) 046801.
\newblock \href {https://doi.org/10.1103/PhysRevLett.128.046801}
  {\path{doi:10.1103/PhysRevLett.128.046801}}.

\bibitem{PhysRevB.67.041301}
Y.~Avishai, et~al., Superconductor-quantum dot-superconductor junction in the
  kondo regime, Phys. Rev. B 67 (2003) 041301.
\newblock \href {https://doi.org/10.1103/PhysRevB.67.041301}
  {\path{doi:10.1103/PhysRevB.67.041301}}.

\bibitem{PhysRevLett.91.266802}
A.~L. Yeyati, et~al., Nonequilibrium dynamics of andreev states in the kondo
  regime, Phys. Rev. Lett. 91 (2003) 266802.
\newblock \href {https://doi.org/10.1103/PhysRevLett.91.266802}
  {\path{doi:10.1103/PhysRevLett.91.266802}}.

\bibitem{Deacon2015}
R.~S. Deacon, et~al., Cooper pair splitting in parallel quantum dot josephson
  junctions, Nature Communications 6 (2015).
\newblock \href {https://doi.org/10.1038/ncomms8446}
  {\path{doi:10.1038/ncomms8446}}.

\bibitem{PhysRevB.29.3035}
P.~Coleman, New approach to the mixed-valence problem, Phys. Rev. B 29 (1984)
  3035--3044.
\newblock \href {https://doi.org/10.1103/PhysRevB.29.3035}
  {\path{doi:10.1103/PhysRevB.29.3035}}.

\bibitem{coleman1985large}
P.~Coleman, Large n as a classical limit $\frac{1}{N}\approx\ni$ of mixed
  valence, J. Magn. Magn. Mater 47~(48) (1985) 323.

\bibitem{PhysRevLett.58.266}
M.~Lavagna, A.~J. Millis, P.~A. Lee, d -wave superconductivity in the
  large-degeneracy limit of the anderson lattice, Phys. Rev. Lett. 58 (1987)
  266--269.
\newblock \href {https://doi.org/10.1103/PhysRevLett.58.266}
  {\path{doi:10.1103/PhysRevLett.58.266}}.

\bibitem{PhysRevLett.57.1362}
G.~Kotliar, A.~E. Ruckenstein, New functional integral approach to strongly
  correlated fermi systems: The gutzwiller approximation as a saddle point,
  Phys. Rev. Lett. 57 (1986) 1362--1365.
\newblock \href {https://doi.org/10.1103/PhysRevLett.57.1362}
  {\path{doi:10.1103/PhysRevLett.57.1362}}.

\bibitem{PhysRevB.59.1637}
P.~Schwab, R.~Raimondi, Andreev tunneling in quantum dots: A slave-boson
  approach, Phys. Rev. B 59 (1999) 1637--1640.
\newblock \href {https://doi.org/10.1103/PhysRevB.59.1637}
  {\path{doi:10.1103/PhysRevB.59.1637}}.

\bibitem{chamoli2022josephson}
T.~Chamoli, Ajay, Josephson transport through parallel double coupled quantum
  dots at infinite-u limit, The European Physical Journal B 95~(9) (2022) 163.
\newblock \href {https://doi.org/https://doi.org/10.1007/s10948-021-06002-w}
  {\path{doi:https://doi.org/10.1007/s10948-021-06002-w}}.

\bibitem{Chamoli2022}
T.~Chamoli, Ajay, Andreev bound states in superconductor–quantum
  dot–superconductor junction at infinite-u limit, Journal of
  Superconductivity and Novel Magnetism 35 (2022).
\newblock \href {https://doi.org/10.1007/s10948-021-06002-w}
  {\path{doi:10.1007/s10948-021-06002-w}}.

\bibitem{PhysRevLett.110.076803}
B.-K. Kim, et~al., Transport measurement of andreev bound states in a
  kondo-correlated quantum dot, Phys. Rev. Lett. 110 (2013) 076803.
\newblock \href {https://doi.org/10.1103/PhysRevLett.110.076803}
  {\path{doi:10.1103/PhysRevLett.110.076803}}.

\bibitem{r62}
J.~O. Island, et~al., Proximity-induced shiba states in a molecular junction,
  Phys. Rev. Lett. 118 (2017) 117001.
\newblock \href {https://doi.org/10.1103/PhysRevLett.118.117001}
  {\path{doi:10.1103/PhysRevLett.118.117001}}.

\bibitem{wendin2007quantum}
G.~Wendin, V.~Shumeiko, Quantum bits with josephson junctions, Low Temperature
  Physics 33~(9) (2007) 724--744.
\newblock \href {https://doi.org/https://doi.org/10.1063/1.2780165}
  {\path{doi:https://doi.org/10.1063/1.2780165}}.

\bibitem{arnault2022dynamical}
Arnault, et~al., Dynamical stabilization of multiplet supercurrents in
  multiterminal josephson junctions, Nano Letters 22~(17) (2022) 7073--7079.
\newblock \href {https://doi.org/https://doi.org/10.1021/acs.nanolett.2c01999}
  {\path{doi:https://doi.org/10.1021/acs.nanolett.2c01999}}.

\bibitem{PhysRevB.98.035438}
G.-Y. Yi, et~al., Suppression of the $0-\pi$ transition in a josephson junction
  with parallel double-quantum-dot barriers, Phys. Rev. B 98 (2018) 035438.
\newblock \href {https://doi.org/10.1103/PhysRevB.98.035438}
  {\path{doi:10.1103/PhysRevB.98.035438}}.

\bibitem{PhysRevResearch.3.033240}
A.~Vekris, et~al., Josephson junctions in double nanowires bridged by in-situ
  deposited superconductors, Phys. Rev. Res. 3 (2021) 033240.
\newblock \href {https://doi.org/10.1103/PhysRevResearch.3.033240}
  {\path{doi:10.1103/PhysRevResearch.3.033240}}.

\bibitem{PRXQuantum.3.030311}
A.~Bargerbos, et~al., Singlet-doublet transitions of a quantum dot josephson
  junction detected in a transmon circuit, PRX Quantum 3 (2022) 030311.
\newblock \href {https://doi.org/10.1103/PRXQuantum.3.030311}
  {\path{doi:10.1103/PRXQuantum.3.030311}}.

\end{thebibliography}

\end{document}